\documentclass[aps, prc, superscriptaddress, twocolumn, floatfix]{revtex4-1}
\usepackage{amsmath, amssymb, amsfonts}
\usepackage{graphicx}
\usepackage{color}
\usepackage{subfig}
\usepackage{grffile}
\usepackage{hyperref}
\graphicspath{{./}}

\newcommand{\figref}[1]{Fig.\,\ref{#1}}

\newcommand{\fm}{\thinspace{\text{fm}}\thinspace}

\begin{document}
\title{What can we learn from Dijet suppression at RHIC?}

\author{C.E.Coleman-Smith} \email{cec24@phy.duke.edu}
\author{B.M\"uller}
\affiliation{Department of Physics, Duke University, Durham, NC 27708-0305}
\date{\today}

\begin{abstract}
We present a systematic study of the dijet suppression at RHIC  using the VNI/BMS parton cascade. We examine the modification of the dijet asymmetry $A_j$ and the within-cone transverse energy distribution (jet-shape) along with partonic fragmentation distributions $z$ and $j_{t}$ in terms of: $\hat{q}$; the path length of leading and sub-leading jets; cuts on the jet energy distributions; jet cone angle and the jet-medium interaction mechanism. We find that $A_j$ is most sensitive to $\hat{q}$ and relatively insensitive to the nature of the jet-medium interaction mechanism. The jet profile is dominated by $\hat{q}$ and the nature of the interaction mechanism. The partonic fragmentation distributions clearly show the jet modification and differentiate between elastic and radiative+elastic modes.
\end{abstract}


\maketitle

\section{Introduction}

Recent measurements showing the strong modification of high energy $E_{t} \sim 100-200$ GeV dijets at the LHC \cite{Aad:2010bu, Chatrchyan:2011sx} in Pb+Pb collisions have sparked an interest in using dijets to quantify jet modification in hot QCD matter. Proposed upgrades to the PHENIX experiment, and continuing development of jet reconstruction at STAR, promise access to dijets in a different kinematic region, $E_{t} \sim 15-65$ GeV at RHIC. The dijet asymmetry 
\begin{equation}
  \label{eqn-aj-define}
A_j = \frac{E_{t,\ell} - E_{t,s}}{E_{t,\ell}+E_{t,s}},
\end{equation}
where $E_{t,\ell}$ is the transverse energy of the leading jet and $E_{t,s}$ is that of the sub-leading jet, has been successfully reproduced by various authors with a variety of models \cite{Qin:2010mn, Young:2011qx, ColemanSmith:2011wd, He:2011pd}. Recent work by Renk \cite{Renk:2012cx} has indicated that these results may be relatively insensitive to the fine details of jet energy loss, suggesting that $A_j$ may be unsuitable for tomographic purposes. If this is indeed the case are there other more differential dijet observables for jet measurements at RHIC? 

With these issues in mind we have used the VNI/BMS parton cascade to  systematically explore dijet suppression at RHIC energy scales under controlled medium conditions. We examine the modification of dijets under variation of the medium radius and temperature, the strong coupling constant and the jet definition in terms of the Anti-Kt cone angle and the leading jet energy cut. 

The VNI/BMS parton cascade model \cite{Geiger:1991nj, Bass:2002fh} provides access to the full jet/medium development at a fixed $\hat{q}$. We run the code in a static uniform-medium mode. Here the medium is modeled as torus of a given radius. The dijet propagation lengths are generated as chords centered on uniformly sampled hard-collision vertices within this torus. The model includes a partonic medium which is treated on an equal footing with the jet. This allows jet partons to escape into the medium and vice-versa. In keeping with the infinite static model of the medium we do not consider hadronization of the jet. As such all results are presented at the partonic level only. Finally, we use anti-kt jet reconstruction throughout to give a somewhat realistic treatment of the jet measurement process.

We begin with a brief description of our model. We then show results for the leading parton energy loss and discuss the modification of dijets at RHIC in terms of: the asymmetry $A_j$, the jet radial profile, and partonic fragmentation distributions $z$ and $j_{t}$.

\section{The Parton Cascade Model}

The parton cascade model (PCM) is a Monte-Carlo implementation of the relativistic Boltzmann transport of quarks and gluons
\begin{equation}
\label{eqn-boltzmann}
p^{\mu} \frac{\partial}{\partial x^{\mu}} F_k(x, p) = \sum_{i}\mathcal{C}_i F_k(x,p).
\end{equation}
The collision term $\mathcal{C}_i$ includes all possible $2\to2$ interactions and final-state radiation $1 \to n$
\begin{align}
  \label{eqn-pcm-collision}
  \mathcal{C}_i F_k(x,\vec{p}) &= \frac{(2 \pi)^4}{2 S_i} \int \prod_{j} d\Gamma_j | \mathcal{M}_i | ^2 \times \notag\\
  &\delta^4\left(P_{in} - P_{out}\right) D(F_k(x, \vec{p})),
\end{align}
$d\Gamma_j$ is the Lorentz invariant phase space for the process $j$, $D$ is the collision flux factor and $S_i$ is a process dependent normalization factor. A geometric interpretation of the total cross-section is used to select pairs of partons for interaction. Between collisions, the partons propagate along straight line trajectories. 

In the VNI/BMS PCM outgoing off-shell partons are brought back on-shell through a medium modified time-like branching. The partons created in this process is subject to a Monte-Carlo LPM effect, see below.

\begin{figure}[htb]
  \includegraphics[width=0.35\textwidth, clip, trim=0.0cm 0.2cm 0.1cm 1.0cm]{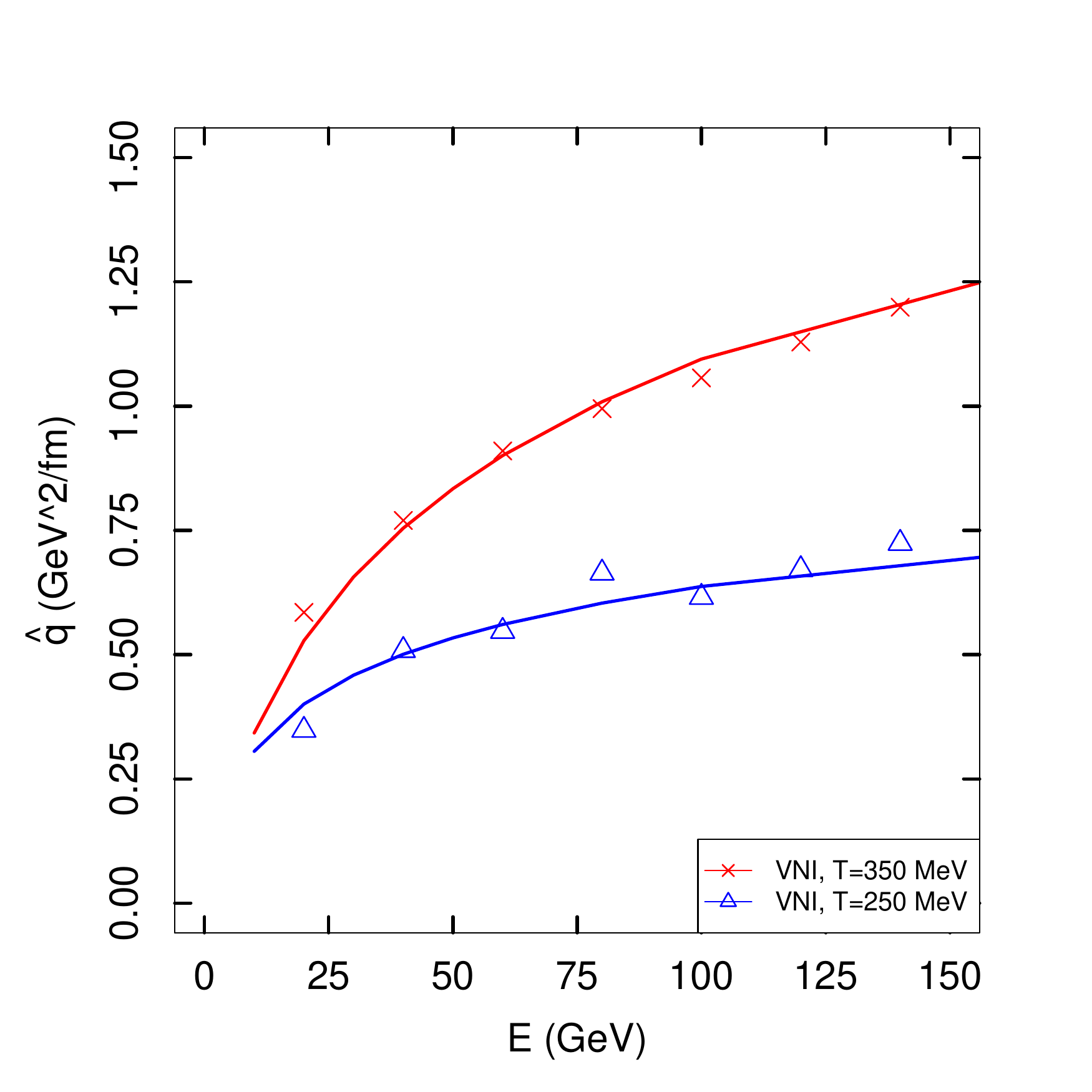} 
  \caption{The transport coefficient $\hat{q}$ for quark jets shown as a function of jet energy and medium temperature, the VNI/BMS data (crosses and open-triangles) agrees well with the theoretical expression (solid lines) \eqref{eqn-qhat}.}
  \label{fig-qhat-temp}
  \vspace{-0.2cm}
\end{figure}

The strong coupling constant for scatterings is nominally held fixed at $\alpha_s = 0.3$ although we shall explore the effects of its variation below. A QGP medium is simulated as a box of thermal quarks and gluons generated at some fixed temperature. Periodic boundary conditions are imposed on the box, whose size is selected to be large enough that if a simulated jet wraps around it will not interact with its own tail. The jet transport coefficient characterizing the amount of transverse momentum acquired by particles in transit through a perturbative thermal medium $\hat{q}$ can be analytically computed \cite{Arnold:2008vd, Shin:2010hu} and also deduced from the simulation. The analytical result is
\begin{align}
  \label{eqn-qhat}
  \hat{q}(T) &= \frac{C_R g^{4} \mathcal{N}(T)}{4\pi} \ln\left(\frac{q_{max}^2(E,T)}{m_D(T)^2} + 1\right), \notag\\
  m_D(T)^2 &= \left(1+ \frac{1}{6}N_f\right) g^{2}T^{2}, \notag \\
  \mathcal{N}(T) &= \frac{\xi(3)}{\xi(2)}\left(1+\frac{1}{4}N_f\right)T^3, \notag \\
  q_{max}(E,T) &= \sqrt{ET}.
\end{align}
where $T$ is the medium temperature $m_D(T)$ is the Debye screening mass, $\mathcal{N}$ is the medium density and $q_{max}$ is a cutoff resulting from the parton scattering kinematics. See \figref{fig-qhat-temp} for a comparison of results from VNI/BMS and \eqref{eqn-qhat}.
The partonic contents of jets created by a suitable event generator and evolved down to $Q_0 = 1$~GeV, in this study PYTHIA 8 \cite{Sjostrand:2006za, Sjostrand08pythia8}, are injected into the medium one at a time, they propagate for a given distance, and their evolution is recorded. The jet evolution is recorded at run-time without reference to any particular jet definition (cone-angle, energy cuts, jet reconstruction algorithm, etc). Instead each particle in the initially inserted jet is marked as being ``jetty'', this jettiness tag is then iteratively applied to all partons that interact with an already jetty-labeled parton. The entire time-evolution of the jet can then be reconstructed using a suitable jet finder (FastJet \cite{Cacciari:2005hq}) and jet definition. In this way the influence of varying jet definitions can be readily studied. We emphasize that the medium partons propagate along with the jet during this process, continually interacting and mixing with the evolving jet. This medium back-reaction is a unique feature of parton cascades.

VNI/BMS is a \emph{simple enough} jet-suppression model. While it may not be suitable for use as a full event generator, lacking hydrodynamical flow and hadronization, the active medium and medium-modified radiation make it a useful test bed for many but not all jet modification features. As we shall show below the effects of hadronization are rather important for certain intra-jet observables. We shall reexamine these observables in a future work with a hadronization solution implemented in the model. 

\section{A Monte-Carlo LPM effect}

The emission of radiation in QCD is not an instantaneous process. There is a time period during which the radiated quanta and the radiator are in a correlated state. If this takes place in a dense medium the correlated wave function may interact with the medium constituents leading to a modification of the decoherence process. If the emitting parton interacts with several additional medium scattering centers then quantum interference effects lead to an in medium path-length dependence for the amount of energy radiated, this is the Landau Pomeranchuk Migdal (LPM) effect \cite{LandauPom, PhysRev.103.1811}.

The formation time of a parton with energy $E$ and time-like virtuality $Q$  
\begin{equation}
  \label{eqn-formation-time-first}
  \tau_f = \frac{E}{Q} \frac{1}{Q} \approx \frac{\omega}{\mathbf{k}_{\perp}^2},
\end{equation}
is the lifetime of this virtual correlated state and characterizes the emission process. 

We implement an approximation to the full pQCD description of the interaction of the radiating system with the medium using a local Monte-Carlo routine in the style of Zapp and Wiedemann \cite{Zapp:2008af}. This prescription reproduces the leading BDMPS-Z \cite{Baier:1994bd, *Baier:1996vi, *Baier:1996kr, Zakharov:1996fv, *Zakharov:1997uu} result for light-parton energy loss in a QGP medium $\Delta E \sim L^2$. This method is particularly appealing since it requires no artificial parametrization of the radiative process, it is a purely probabilistic medium-induced modification.

In VNI/BMS a set of partons is produced by time-like branching after an inelastic scattering. The branching process evolves the initial time-like virtuality $Q$  of the outgoing parton pair downwards to a cutoff $Q_0 \simeq 1$~GeV, converting virtuality into energy and transverse momentum of radiated partons. The formation time of these evolving partons \eqref{eqn-formation-time} is computed for each intermediate stage of the branching process. The hardest gluon in the final state of the time-like branching is selected to be the probe parton. 

The formation times 
\begin{equation}
\label{eqn-formation-time}
\tau_f^{0} = \sum_{\mbox{branchings}}\frac{\omega}{\mathbf{k}_{\perp}^2},
\end{equation}
 for all sub-branchings that occur during the time-like branching procedure leading up to the production of the probe gluon are summed, giving the initial formation time for the probe gluon. The selection of the hardest gluon to represent the shower reflects the dominance of gluon re-scattering in QCD interference processes. 

The probe parton is allowed to propagate through the medium and interact elastically with other partons in the medium. After the n-th scattering its formation time is recalculated as
 \begin{equation}
   \label{eqn-formation-time-recalc}
   \tau_f^{n} = \frac{\omega}{\left(\mathbf{k}_{\perp} + \sum_{i=1}^{n} \mathbf{q}_{\perp,i}\right)^2}. 
 \end{equation}
This tends to decrease the formation time relative to the original value with each additional scattering, leading to an increased energy loss rate. This simulates the emission of the shower with coherent interference from $n$ centers. After the formation time expires, the radiation is considered to have decohered from the initiating particles and all involved partons may interact and radiate.

\begin{figure}[ht]
    \includegraphics[width=0.5\textwidth]{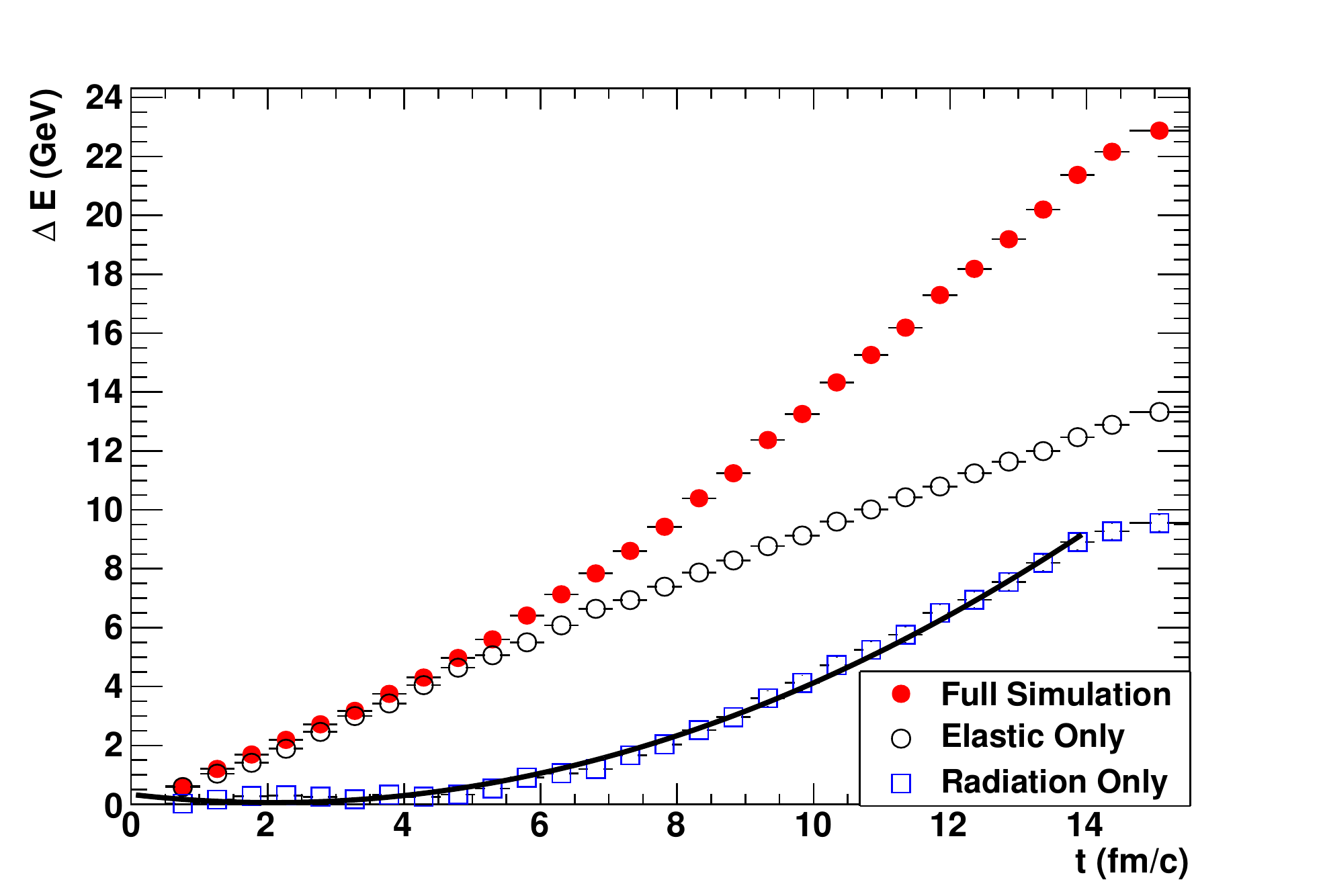} 
    \caption{The total energy loss (red closed circles) for the leading parton in a $100$~GeV jet in a medium at $T=0.35$~MeV, the contributions from elastic and radiative processes (black open circles and blue open boxes) are shown. A fit to \eqref{eqn-bdmps-z} is shown as the black solid line.}
  \label{fig:detotal}
  \vspace{-0.1cm}
\end{figure}

The performance of this prescription has been verified by considering the energy loss of a $100$~GeV jet inserted into a medium at fixed temperature $T = 0.35$~GeV, see \figref{fig:detotal}. The elastic energy loss in the VNI/BMS parton cascade has been previously shown to agree well with perturbative calculations \cite{Shin:2010hu}. The measured radiative energy loss is well described by BDMPS-Z \cite{Baier:1996kr} which gives,
\begin{equation}
  \label{eqn-bdmps-z}
  -\Delta E = \frac{\alpha_s C_{R}}{8} \frac{\mu^{2}}{\lambda_g} L^2 \log \frac{L}{\lambda_g}.
\end{equation}

By fitting the radiation only curve (the blue open squares in \figref{fig:detotal}) with this functional form we match the leading coefficient to within $20\%$.

The elastic energy loss process is dominant for short times since the average formation time arising from the first hard collision of the leading-parton with the medium is $\langle \tau_f \rangle \sim 5$~\fm/c. This leads to the initial suppression of the radiative process as we have required that no energy  be lost from the radiating parton until the formation times of the radiated quanta have expired.

\section{Results}

We generated candidate dijet events from Monte-Carlo pp collisions at $\sqrt{s} = 200$~GeV using PYTHIA 8. Anti-kt reconstruction was used to identify satisfactory dijet events \cite{Cacciari:2008gp}. The following kinematic requirements were always imposed: $p_{t,min} = 5$ GeV, $\Delta |\eta_{12} | < 1.1$, $\Delta \phi_{12} > \pi / 2$ where $\Delta \eta_{12}$ and $\Delta \phi_{12}$ are measured between the two reconstructed jets in the event. The constituent partons of each identified jet were translated from PYTHIA into the parton cascade medium for modification. The two jets in each event are evolved separately.  Table \ref{table-factors} shows the ranges of each parameter  we included in the study. The study contains a total of 16000 dijet events for each combination of parameters. For a given medium radius $R_{med}$ we generate dijet path lengths by uniformly sampling $2d$ jet vertices within a circle of radius $R_{med}$ and then generating a chord with a uniform angular distribution. The two chord segments define the path lengths for the dijet pair. The leading jet is defined as the jet with the highest energy at the point of observation, the leading jet from the vacuum event may not be the leading jet after medium evolution.

\begin{table}
  \begin{tabular}{l | r}
    parameter & range of values \\
    \hline 
    jet cone angle $R$ & 0.2, 0.3, 0.4 \\
    lead energy cut $E_{t,\ell}$ & 25, 35, 50 GeV \\
    interaction mechanism & elastic, elastic \& radiation\\
    Temperature $T$ & 250, 350, 450  MeV \\
    strong coupling $\alpha_s$ & 0.25, 0.3, 0.35, 0.4, 0.6\\
    medium radius $R_{med}$ & 1, 3, 5 \fm
  \end{tabular}
  \caption{16000 dijet events were generated for each combination of the above factors}
  \label{table-factors}
  \vspace{-0.5cm}
\end{table}
We consider the dijet asymmetry $A_j$, the partonic jet radial profile and the partonic distribution of energy within the jet in terms of the variables $z = p_{t}/E_t \cos( \Delta r)$ and $j_{t} = p_{t} \sin (\Delta r)$, where $p_t$ is the transverse momentum of a parton within the jet, $E_t$ is the total transverse momentum of the jet and $\Delta r = \sqrt{(\Delta \phi^2 + \Delta \eta^2)}$ is the angle between this parton and the jet axis.

\begin{figure}[ht]
  \includegraphics[width=0.45\textwidth]{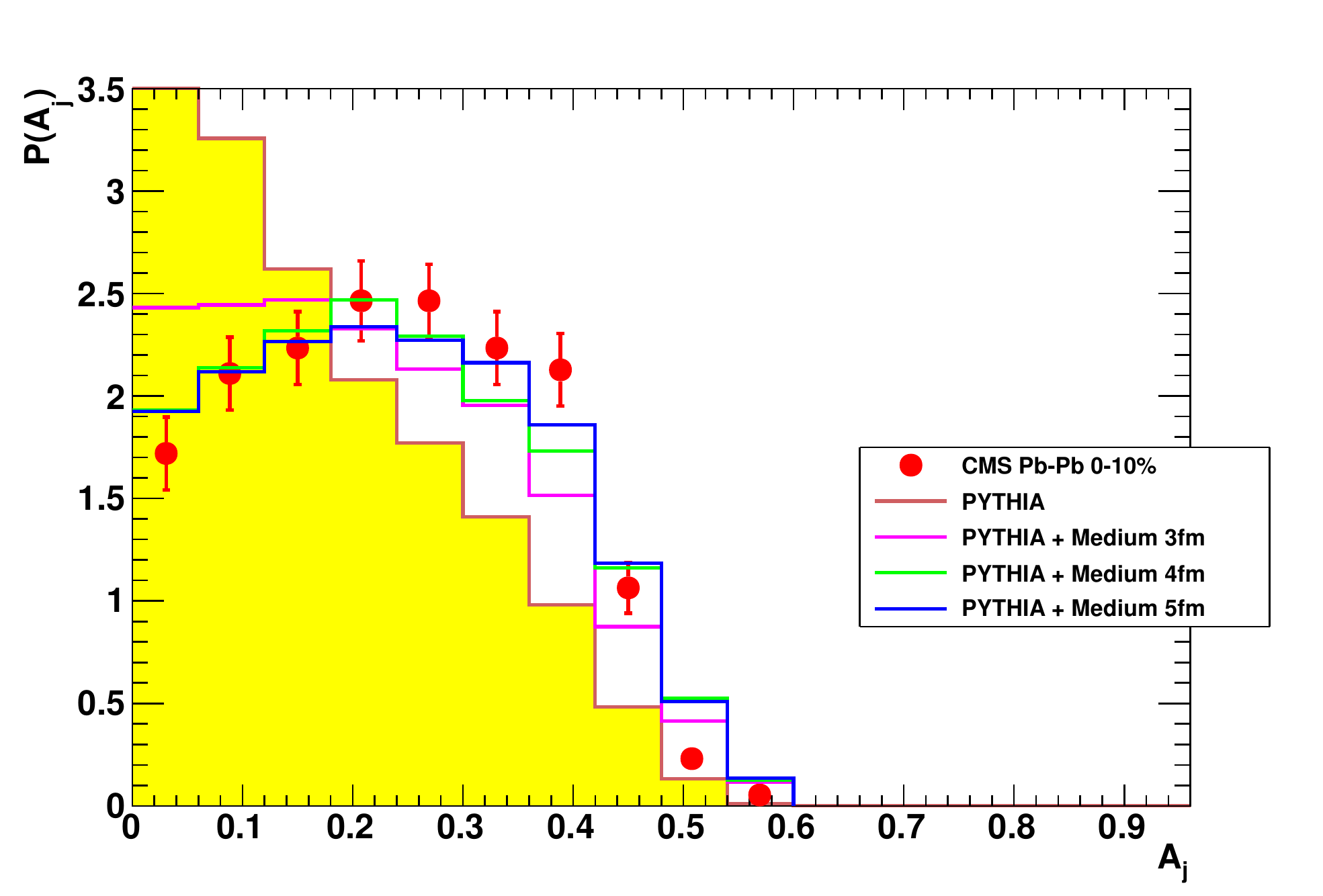}
  \caption{ The VNI/BMS post diction of the dijet asymmetry as measured at CMS, jets are reconstructed with Anti-Kt $R=0.5$ and a smearing $E_t^{\star} \sim~ \mbox{N}(E_t, 1.2\sqrt{E_t})$ is applied to simulate the detector response.}
  \label{fig-aj-cms}
\end{figure}

We have previously applied VNI/BMS \cite{ColemanSmith:2011wd} with a static medium to jets at LHC scales $\sqrt{s} = 2.76$~ATeV, see \figref{fig-aj-cms}. Here we followed the dijet kinematics as described by CMS \cite{Chatrchyan:2011sx}, $E_{t,\ell} > 120$~GeV, $E_{t,s} > 50$~GeV and $\Delta \phi_{12} > 2 \pi / 3$. A Gaussian smearing was applied to reconstructed jet energies to approximate detector response. In this case a medium with $T=350$~MeV and $R_{med} = 3$~fm gives a relatively good fit to the CMS results \cite{Chatrchyan:2011sx}. This suggests that VNI/BMS can produce reasonable jet modification. Given the complexity of the actual detector response we present all further results {\emph without} any attempt at smearing with the hope that particular detector responses can be folded into the data as required.

\begin{figure*}[p]
  \includegraphics[width=0.4\textwidth, clip, trim=0.2cm 0.1cm 0.5cm 0.5cm]{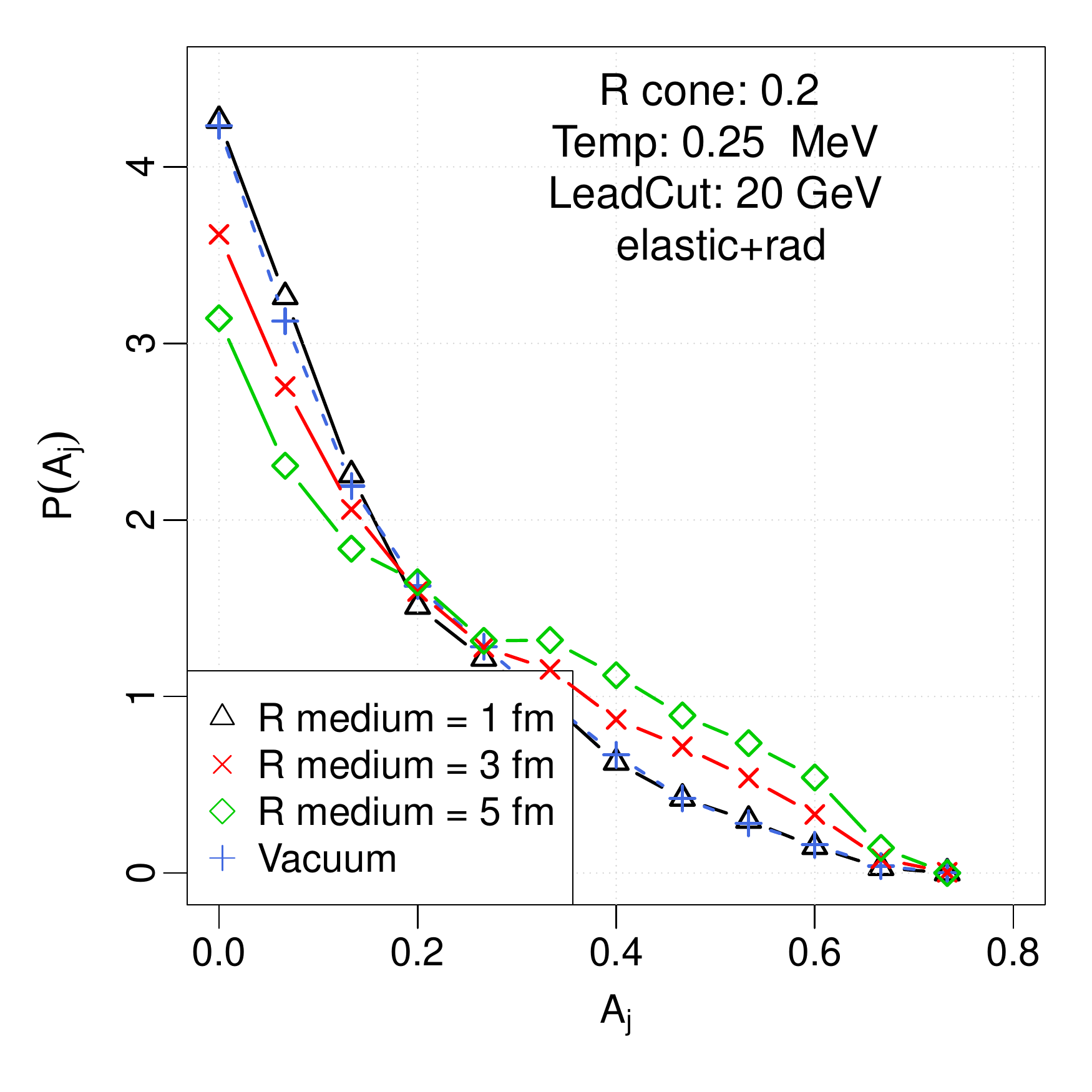}
  \includegraphics[width=0.4\textwidth, clip, trim=0.2cm 0.1cm 0.5cm 0.5cm]{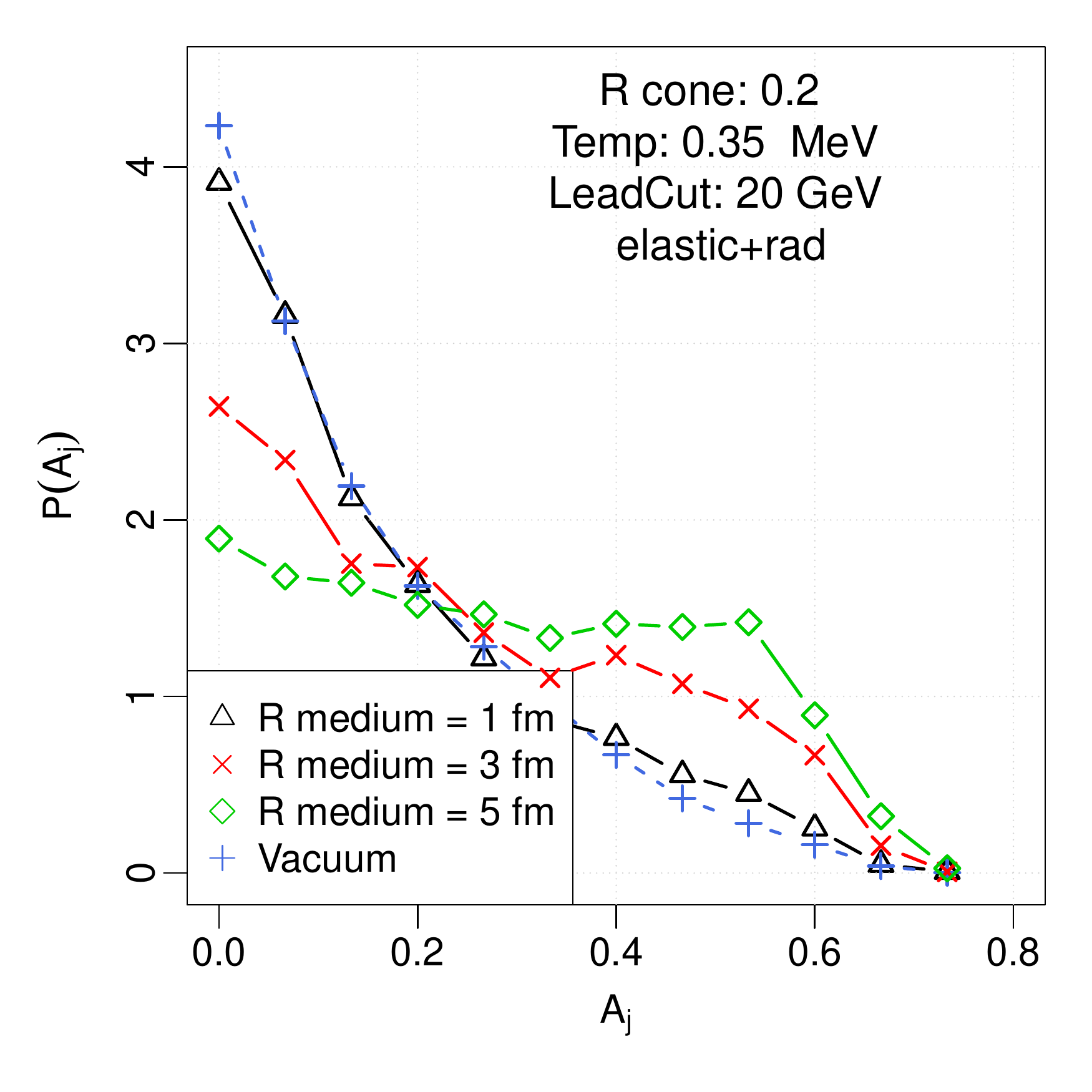}
  \caption{ The dijet asymmetry as a function of the path length of the leading and sub-leading jets,  for a medium with $T=250$ MeV (left) and  $T=350$ MeV (right). Both figures show anti-kt reconstructed jets of radius $R=0.2$ with $E_{t,\ell} \ge 20$ GeV and all events include both radiative and elastic energy loss.}
  \label{fig-aj-z}
\end{figure*}

We now consider the modification of the RHIC dijet asymmetry under variation of all the factors in our design.  In \figref{fig-aj-z} we show the variation of $A_j$ with the medium radius for two different medium temperatures. The higher temperature has a $\hat{q}$ roughly double the lower, see \figref{fig-qhat-temp}. The vacuum jet distribution falls off very quickly, jets with an initial $E_{t,\ell} > 65$~GeV are exceedingly rare. Increasing the medium temperature from $250$~MeV to $350$~MeV leads to a depletion of jets with a small modification shifting the dijet distribution towards higher $A_j$ values.  By applying an energy cut to differentiate the leading and sub-leading jets in a dijet pair we have biased our results. We select for leading jets which travel a short distance, and are less modified,  along with sub-leading jets that travel a very long distance and so experience a larger integrated $\hat{q}$. This surface bias gives a large contribution to $A_j$ relative to the vacuum result. For the remainder of the analysis we have fixed $R_{med} = 5$~\fm.

\begin{figure*}[p!]
  \includegraphics[width=0.45\textwidth]{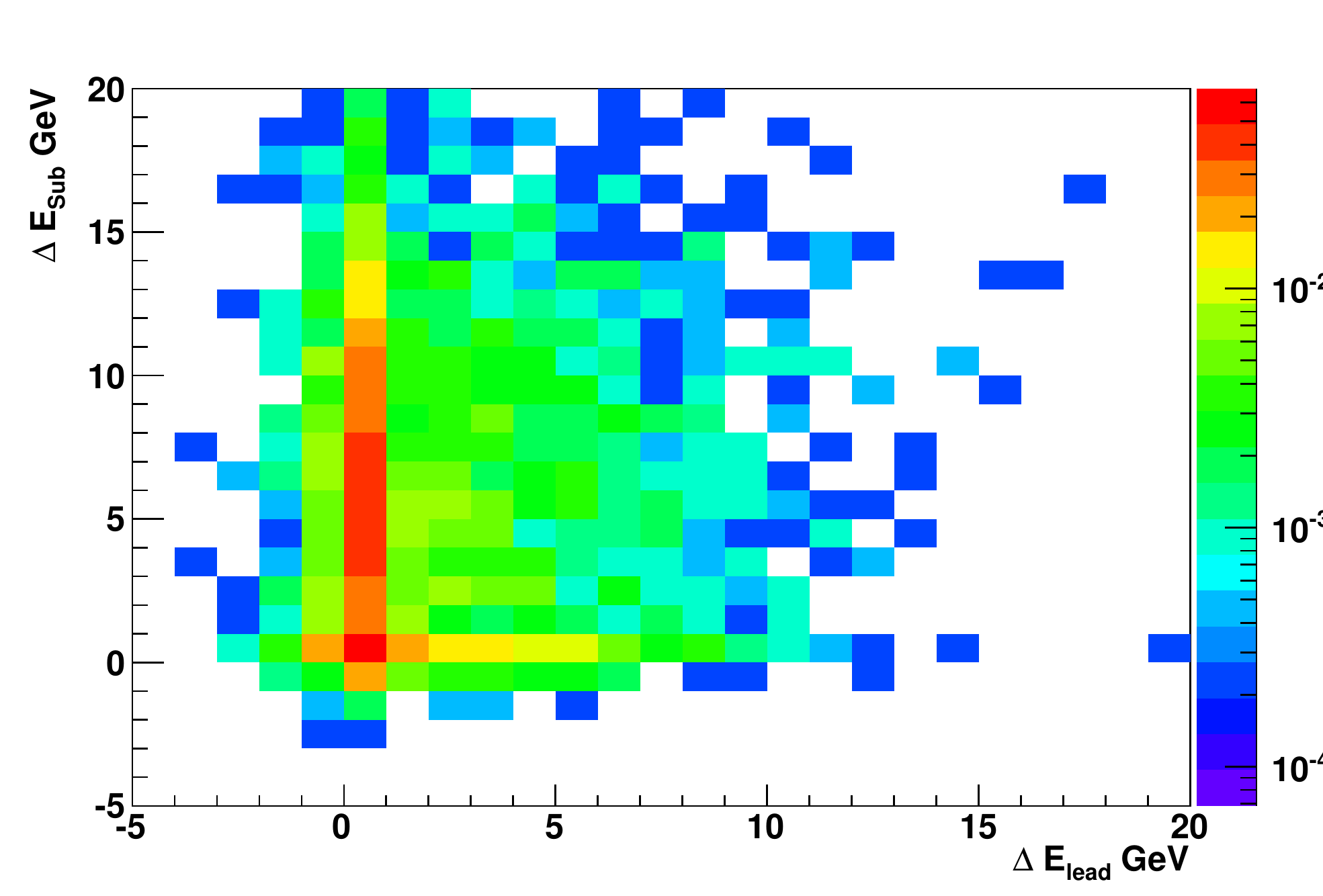}
  \includegraphics[width=0.45\textwidth]{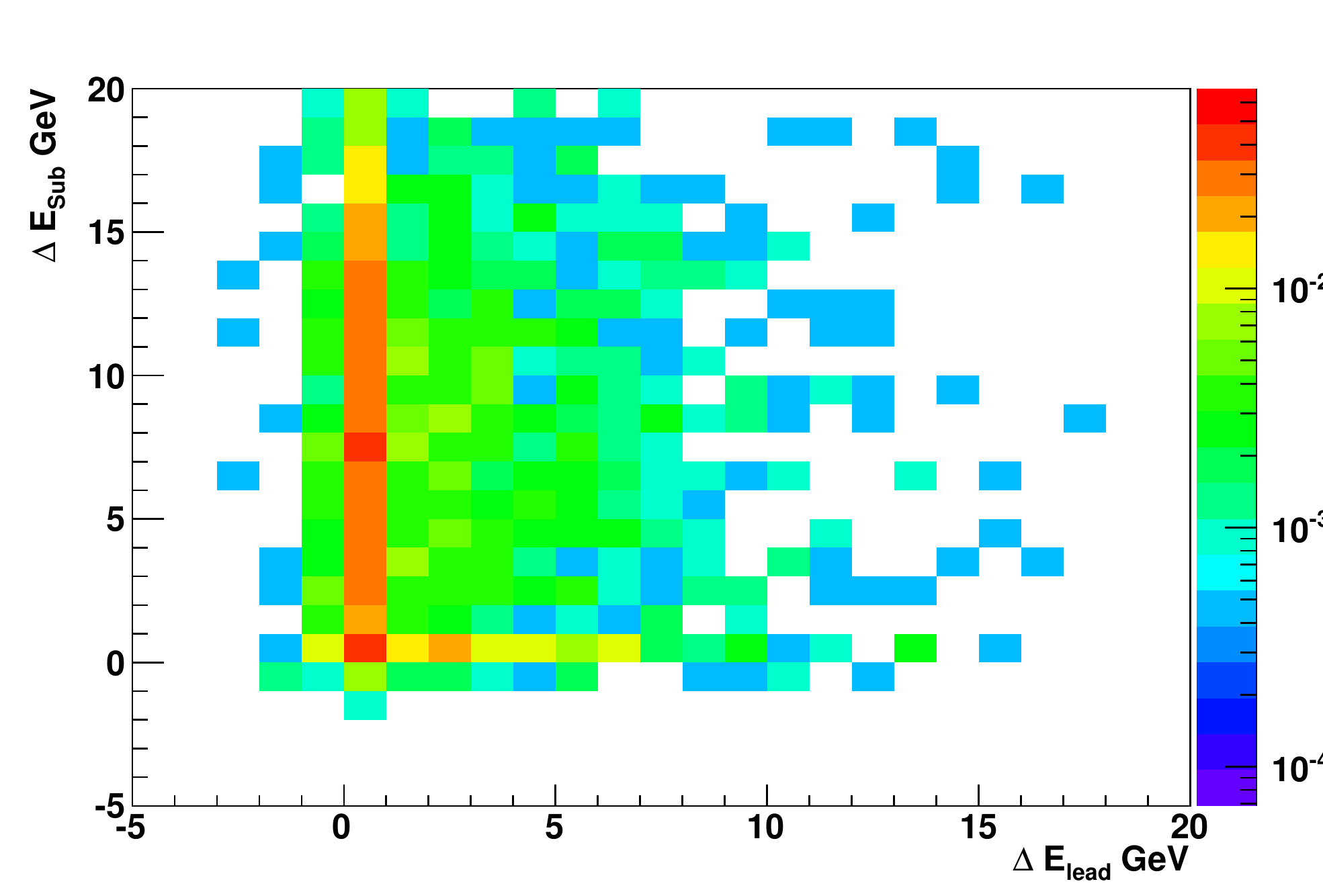}
  \caption{The joint energy loss distributions for leading and sub-leading jets at $T_{med}=350$~MeV with $R_{med} = 5$~\fm and anti-kt $R = 0.2$. The left figure shows the elastic only mode, right figure shows elastic + radiation. }
  \label{fig-de-joint}
\end{figure*}

\begin{figure*}[p!]
  \includegraphics[width=0.45\textwidth]{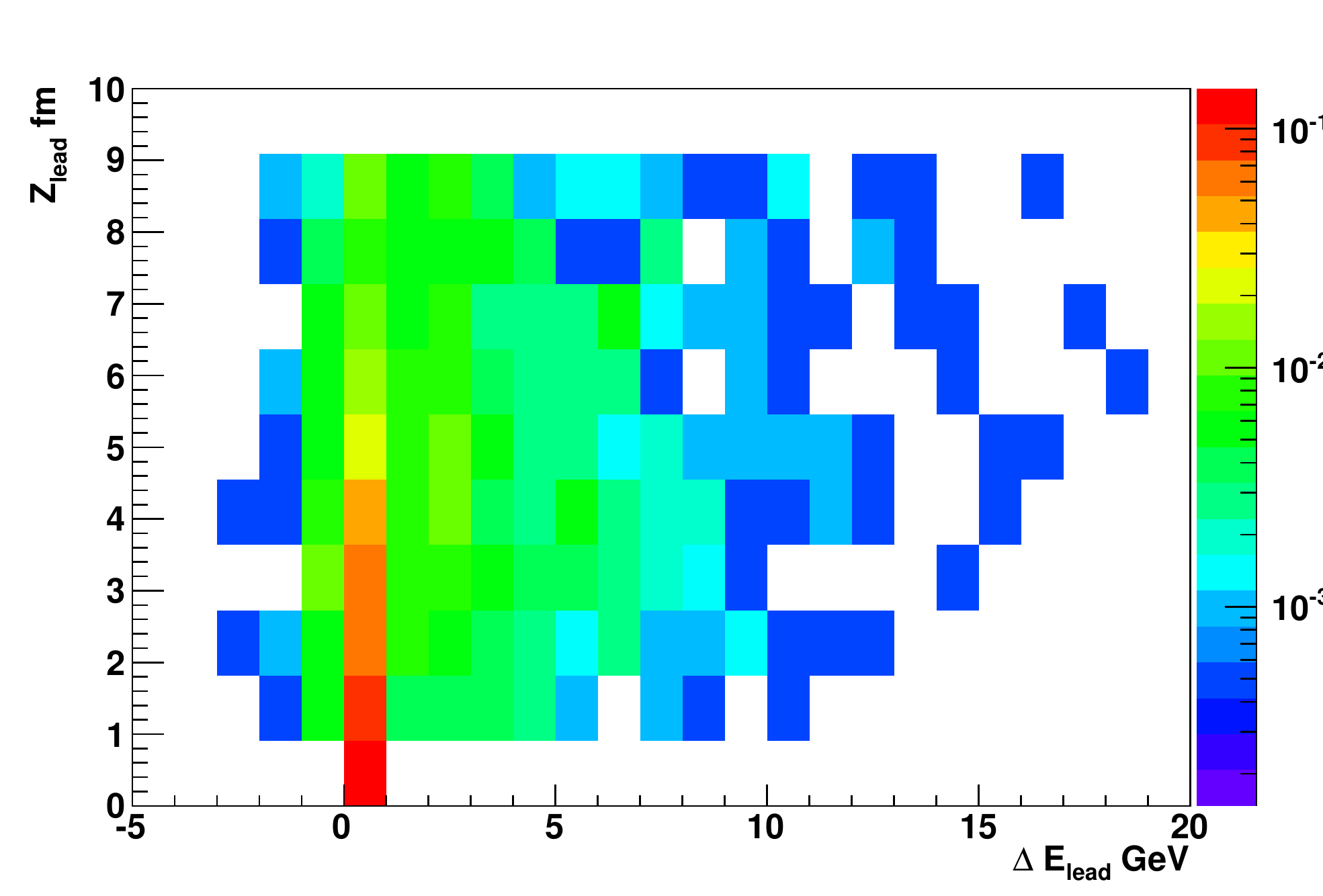}
  \includegraphics[width=0.45\textwidth]{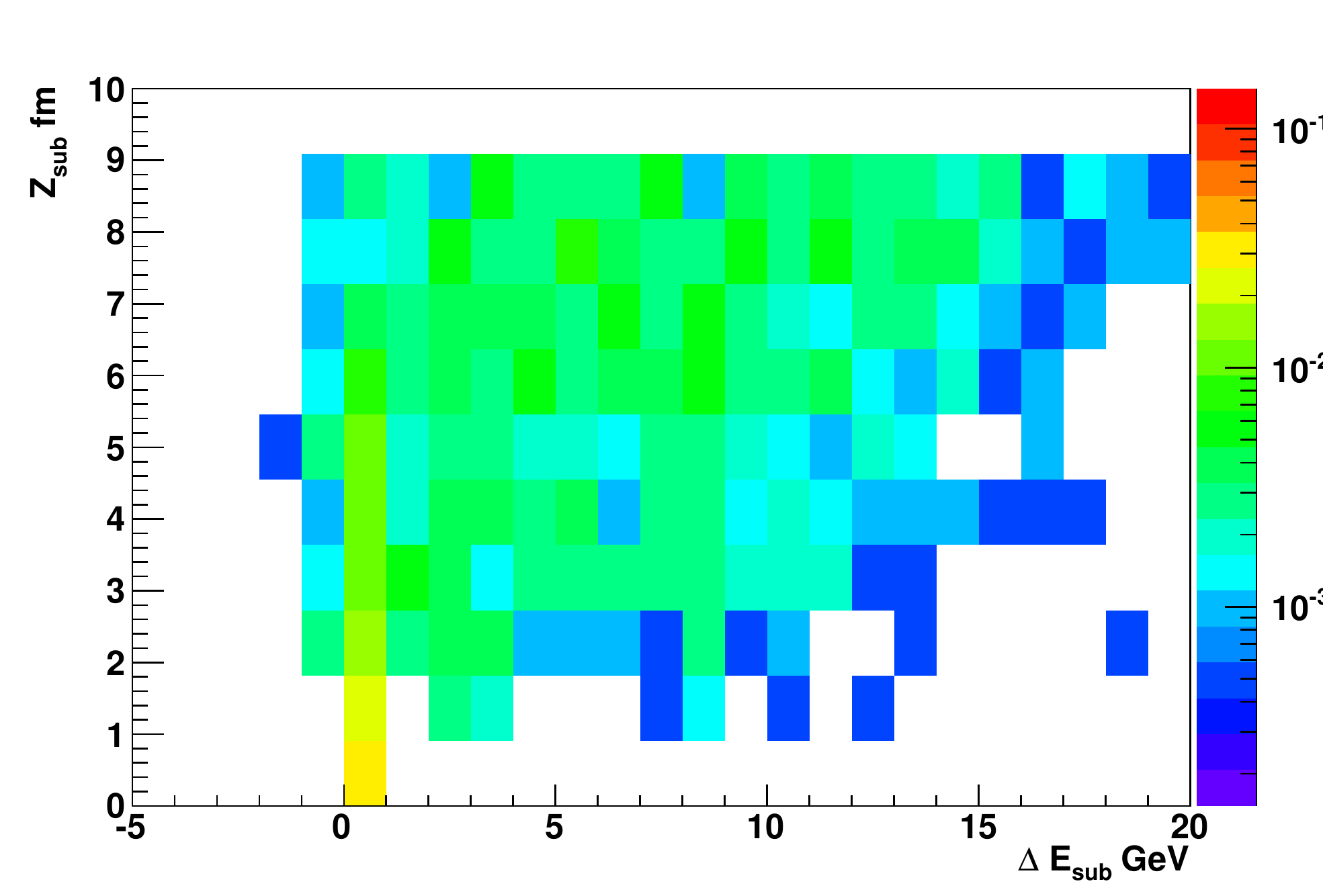}
  \caption{Showing the distribution of energy loss against path length for leading (left) and sub-leading (right) jets at $T_{med}=350$~MeV with $R_{med} = 5$~\fm and anti-kt $R = 0.2$ in the full radiation+elastic simulation.}
  \label{fig-de-path}
\end{figure*}

The joint leading and sub-leading energy loss distribution for elastic only interactions and the full simulation is shown in \figref{fig-de-joint} for jets with $T_{med} = 350$~MeV and $E_{t,\ell} > 20$ GeV. In both cases the distribution is peaked at zero, most jets simply don't lose any energy which contributes to the relatively large peaks at $A_j \sim 0$. The leading partons may lose energy by interacting with the medium this energy has to be transported outside of the jet cone for the $E_t$ of the jet to be modified. The line $\Delta E_{t,\ell} = \Delta E_{t,s}$ is also responsible for the peak at small $A_j$.  The leading jet rarely loses much energy in either scheme while the sub-leading jet is noticeably more modified in the radiative scheme. In \figref{fig-de-path} the distribution of energy loss against distance traveled is shown. The leading jet shows a very strong surface bias. The leading jets travel a few fm and lose no energy. The sub-leading jet distribution is also peaked at $z=0$ but this peak is smaller and the bulk of the distribution is spread into a wide range of path lengths and energy losses. An understanding of the path length dependence of the leading and sub-leading jet energy loss is vital for understanding the tomographic implications of $A_j$. 

\begin{figure*}[p]
      \includegraphics[width=0.45\textwidth, clip, trim=0.2cm 0.1cm 0.5cm 1cm]{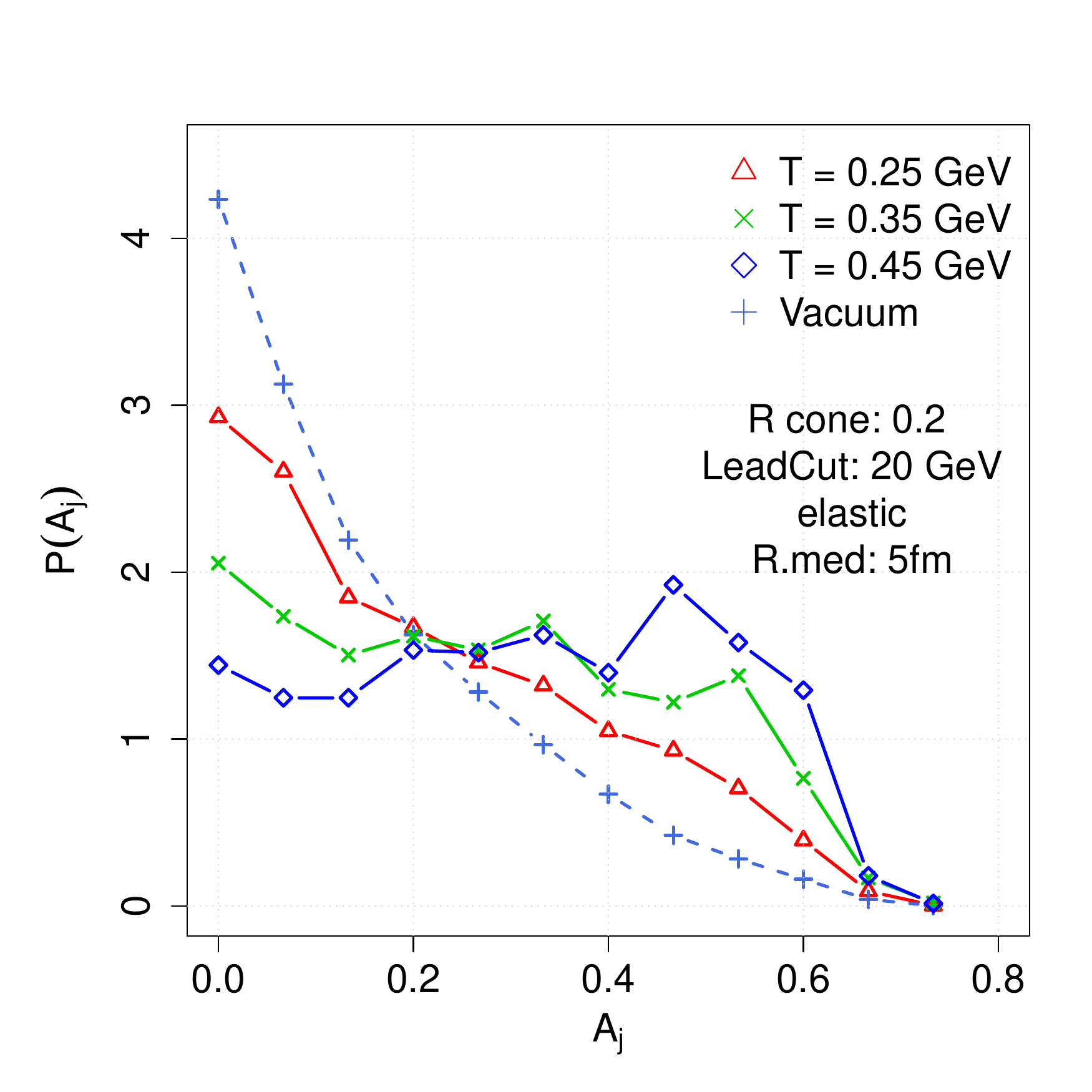}
      \includegraphics[width=0.45\textwidth, clip, trim=0.2cm 0.1cm 0.5cm 1cm]{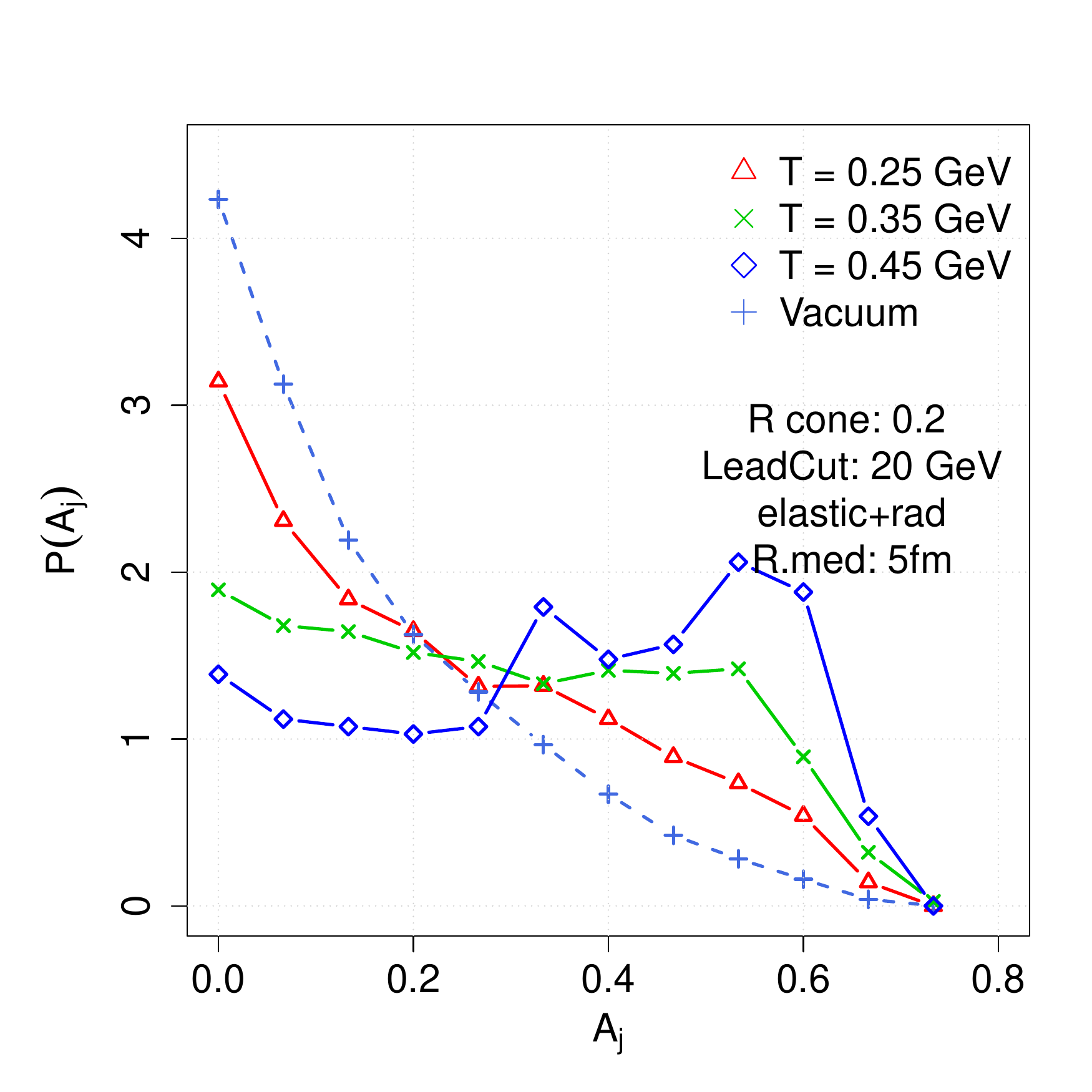}
  \caption{Showing the influence of the medium temperature and the jet-medium interaction mechanism for dijets with $E_{t1} \ge 20$ GeV. The left panel shows elastic scattering only, the right panel shows the full elastic and radiative process. All events are for anti-kt reconstructed jets with $R=0.2$ and $R_{med}=5$~\fm. }
  \label{fig-aj-temp}
\end{figure*}

Since $\hat{q} \propto T^3$ the dijet asymmetry depends strongly upon the medium temperature. In \figref{fig-aj-temp} we show results of varying the medium temperature. Increasing medium temperature and therefore $\hat{q}$ leads to increased jet-medium interactions and a strong swing in the observed $A_j$. The difference between the elastic only results and the full simulation including radiation is remarkably small at lower temperatures. The modification of the radiative jets is somewhat greater at $T=0.45$~GeV. For jets at these scales the asymmetry is relatively insensitive to the details of the jet interactions.

\begin{figure*}[p]
  \includegraphics[width=0.45\textwidth]{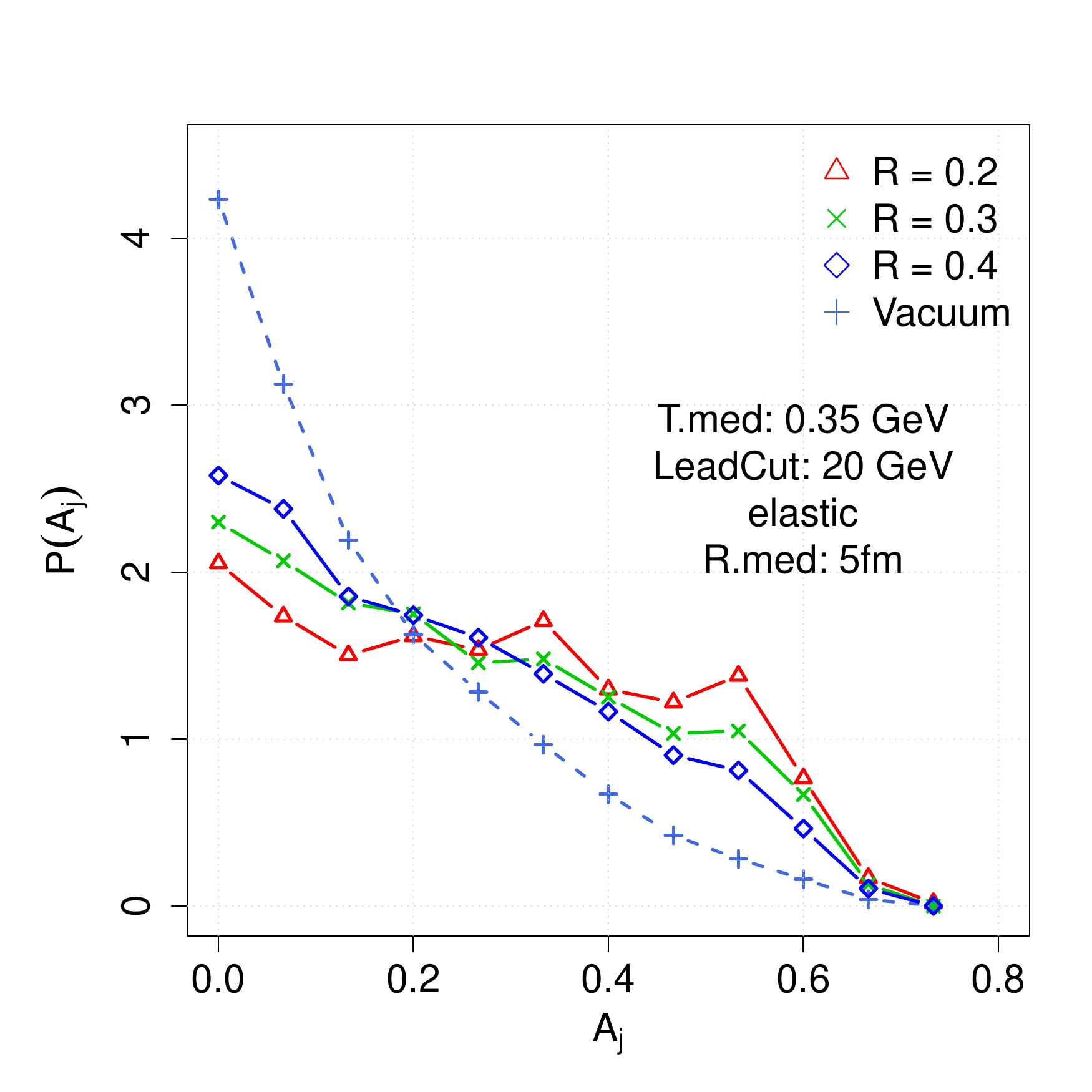}
  \includegraphics[width=0.45\textwidth]{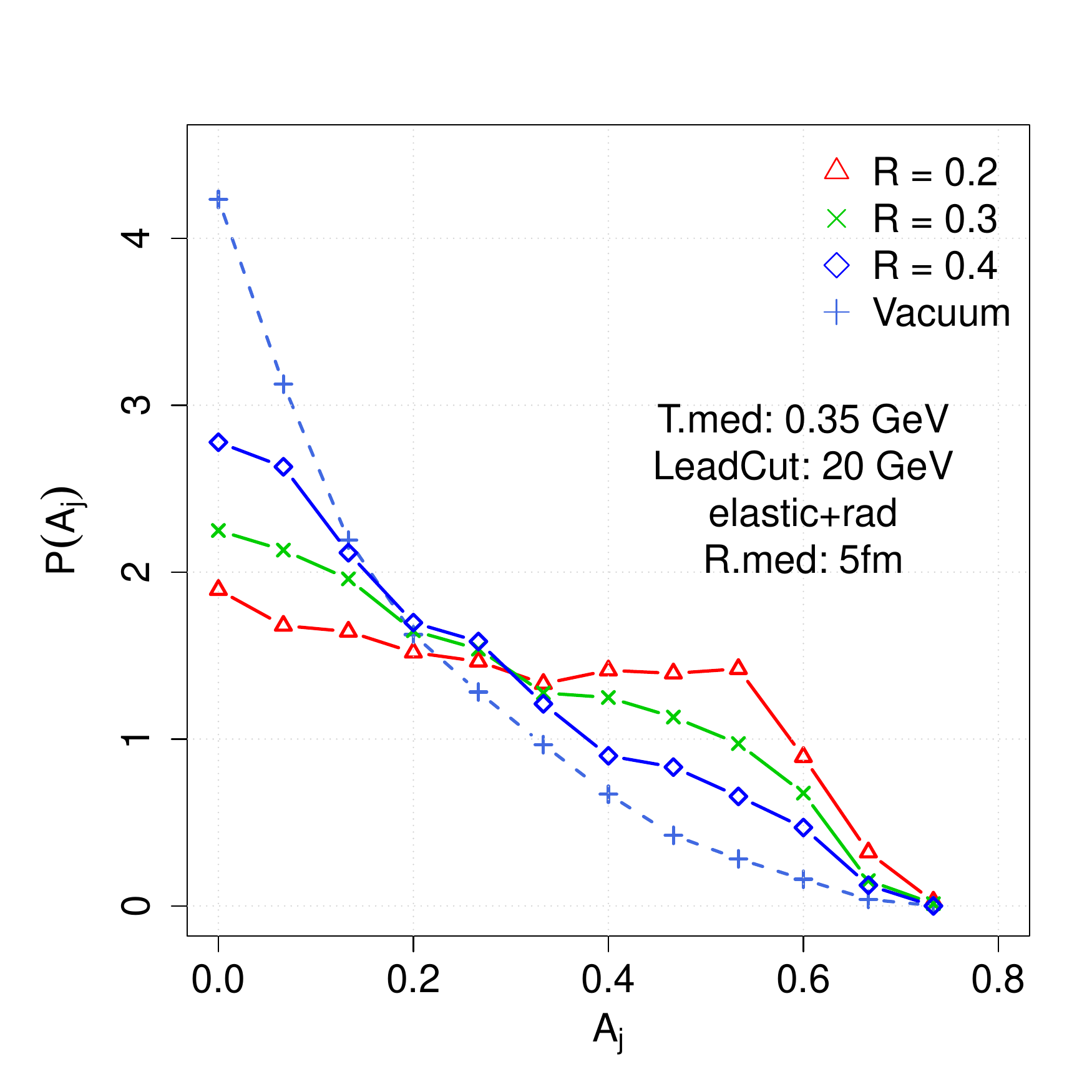}
  \caption{Showing the variation of $A_j$ under variation of $R$ the Anti-Kt jet cone radius, elastic only (left) and elastic + rad (right). The medium radius is kept fixed at $R_{med} = 5$~\fm with $T=0.35$~GeV and $E_{t,\ell} > 20$ GeV. }
  \label{fig-aj-R}
\end{figure*}

In \figref{fig-aj-R} we show $A_j$ as a function of the anti-kt cone angle $R$ for jets in mediums with $T=250$~MeV and $T=350$~MeV. As $R$ is increased the amount of dijet modification is significantly reduced. It is important to note that the only partons which can be included as part of the measured jets were either directly created by the jet or are those which have scattered with jet partons. The thermal medium is currently artificially excluded from the jet finder, this removes uncertainty associated with background removal. As $R$ increases more of the relatively soft radiated partons and forward scattered medium partons are included in the jet definition along with original hard core. This leads to higher reconstructed jet energies at larger $R$'s which in turn leads to the observed reduced $A_j$. The effect is proportional to the distance traveled by the sub-leading jet, jets which have traveled shorter distances have built up less of a cloud of soft partons and picked up fewer medium partons by elastic forward scattering. The rate of transverse diffusion of these soft partons is proportional to $\hat{q}$, so the jet-cone/cloud will be wider at higher medium temperatures. 

\begin{figure*}[p]
    \includegraphics[width=0.45\textwidth]{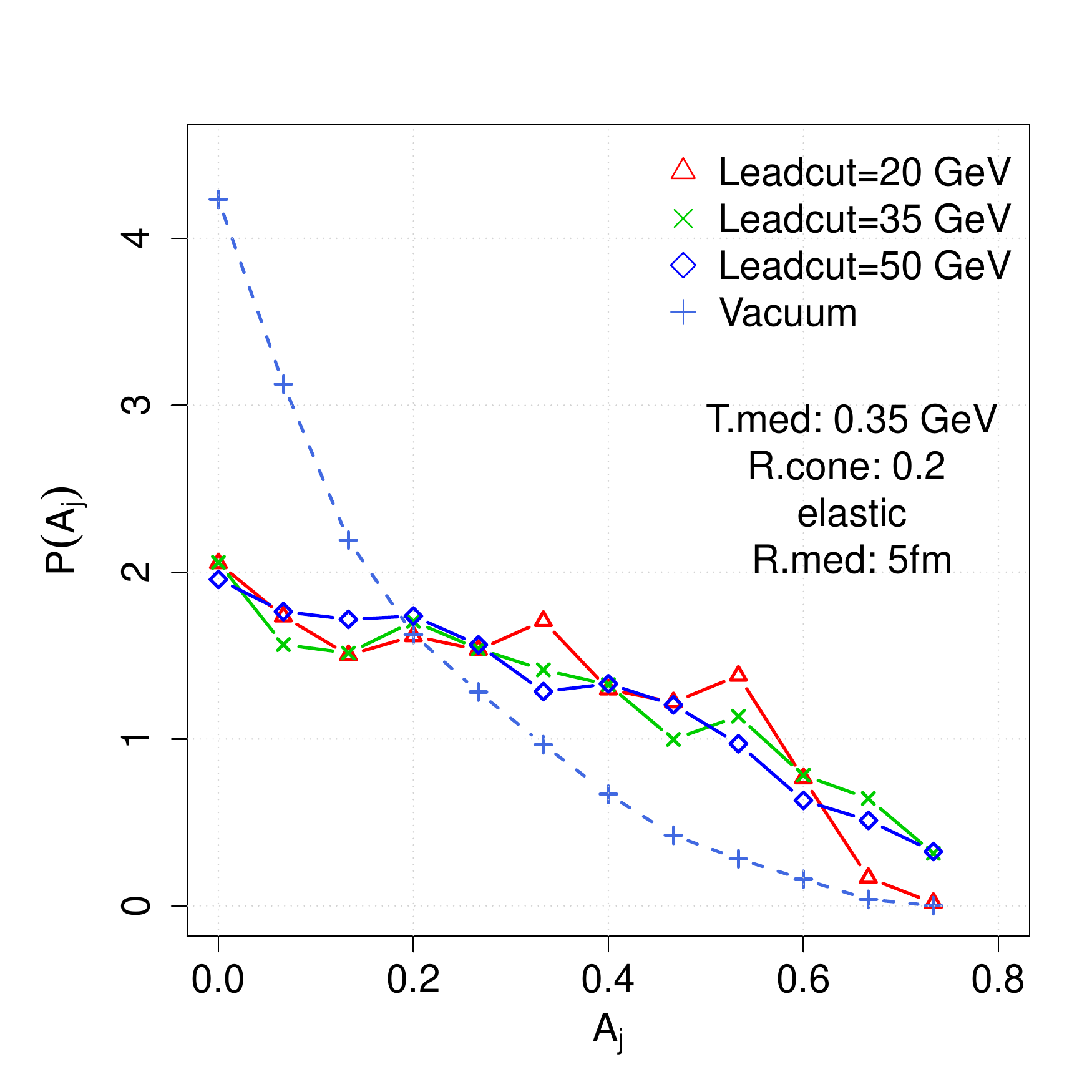}
    \includegraphics[width=0.45\textwidth]{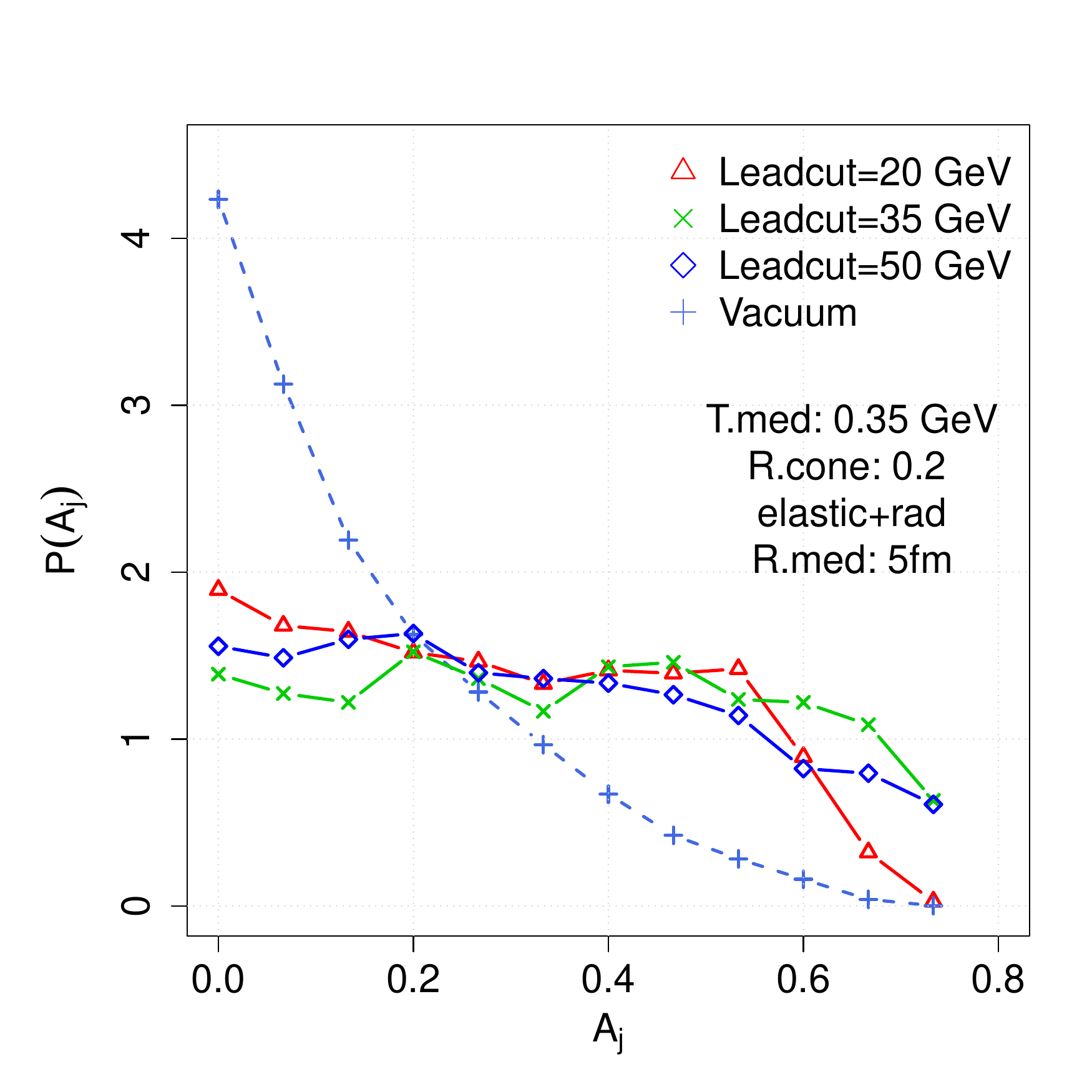}
  \caption{ The influence of applying cuts to the leading jet energy on $A_j$. The left panel shows only elastic scattering processes, the right panel shows radiation and elastic scattering. All events shown are for anti-kt reconstructed jets with $R=0.2$ and $R_{med} = 5$~\fm.} 
  \label{fig-aj-cut}
\end{figure*}

\begin{figure*}[p]
  \includegraphics[width=0.45\textwidth]{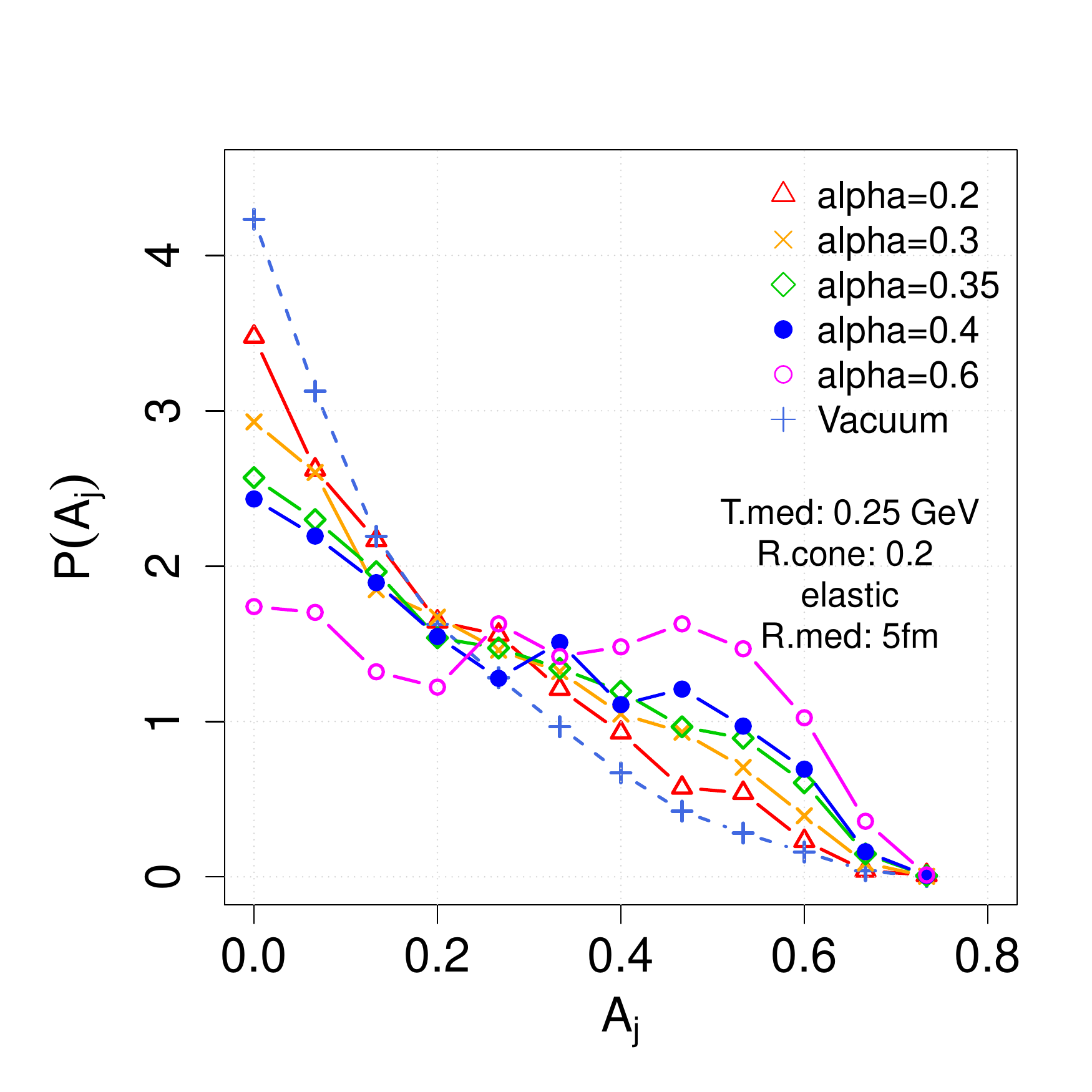}
  \includegraphics[width=0.45\textwidth]{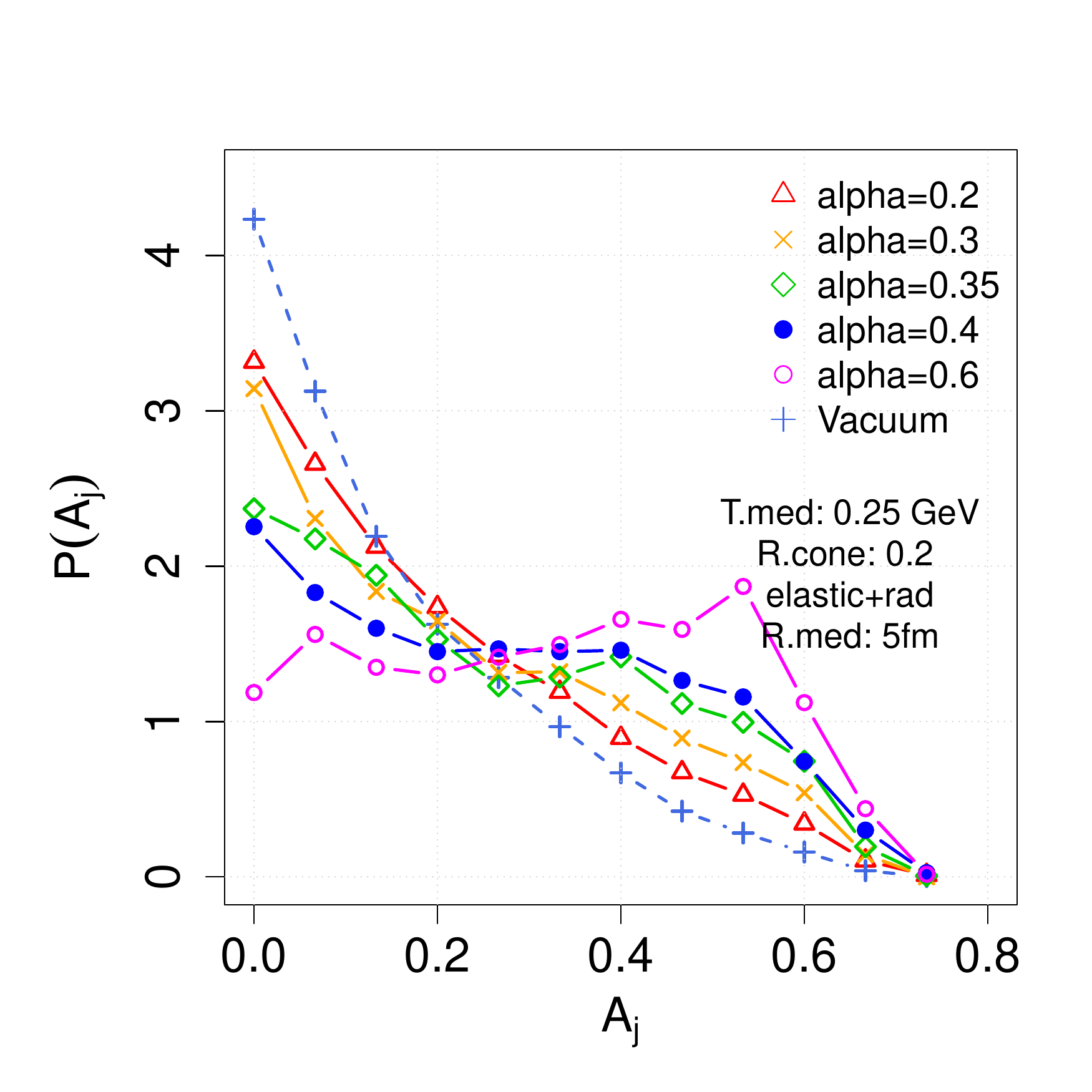}
  \caption{Variation of the coupling constant $\alpha_s$,  elastic only (left) and elastic + rad (right), for jets in a medium at $T=250$~MeV reconstructed with anti-kt $R=0.2$.}
  \label{fig-alpha-vary}
\end{figure*}

The leading jet energy cut is varied in \figref{fig-aj-cut}, at fixed $T=350$~MeV and $R=0.2$, the dijet asymmetry appears to be relatively insensitive to this parameter. We explore the variation of the strong coupling constant $\alpha_s$ in \figref{fig-alpha-vary}. The strength of elastic interactions are scales with the strong coupling and so $\hat{q} \propto \alpha_s^2$. The radiative process itself also depends upon the strong coupling, the total cross section for emission in the DLA (double log approximation) scheme is $\sigma_N \propto (g_{s}^2 log^2)^N$ where the large logs arise from the kinematics of small angle radiation \cite{Dokshitzer:1991wu}. 
Given these considerations it is interesting to note that the relative similarity of the results with and without radiation. 

\begin{figure*}[p]
  \includegraphics[width=0.45\textwidth]{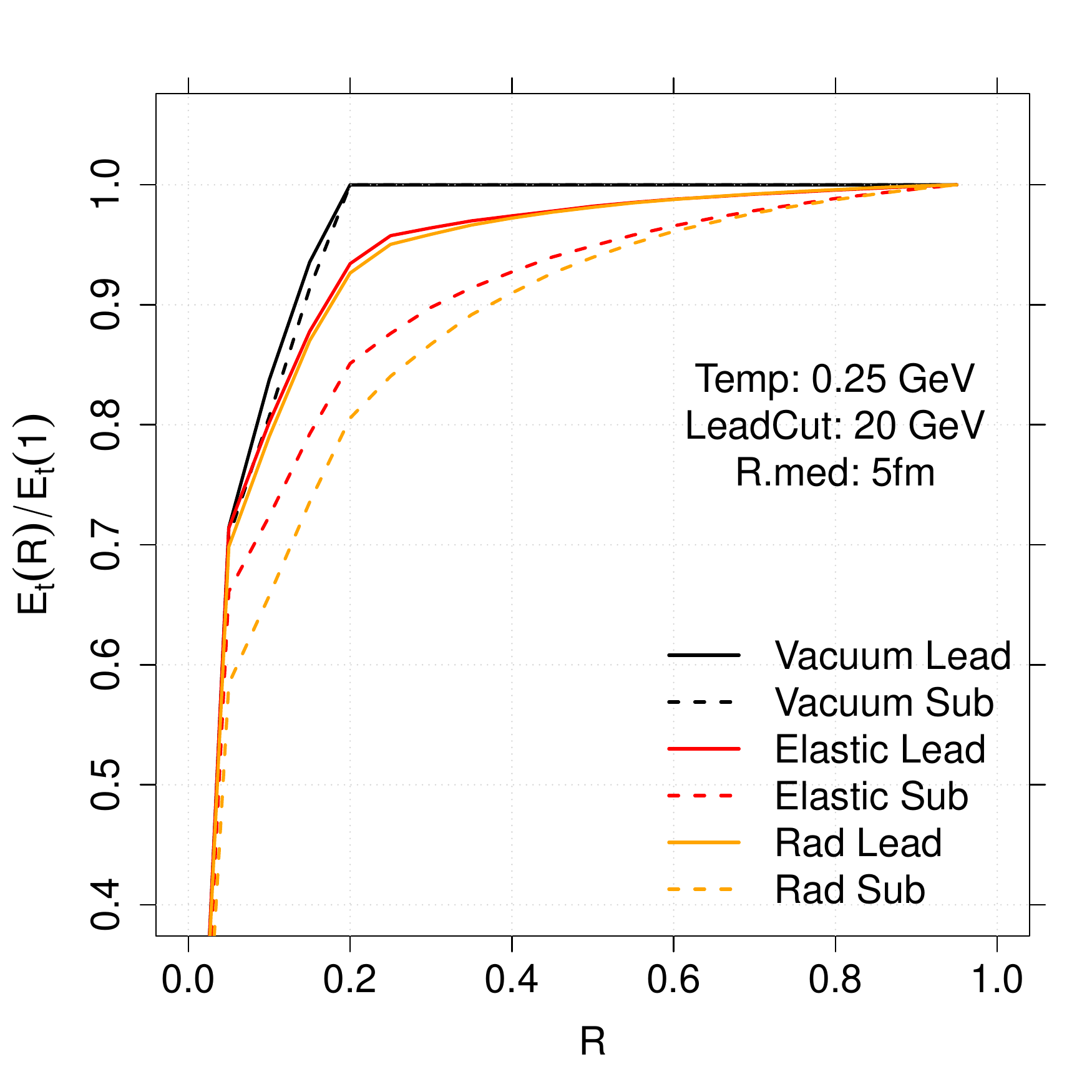}
  \includegraphics[width=0.45\textwidth]{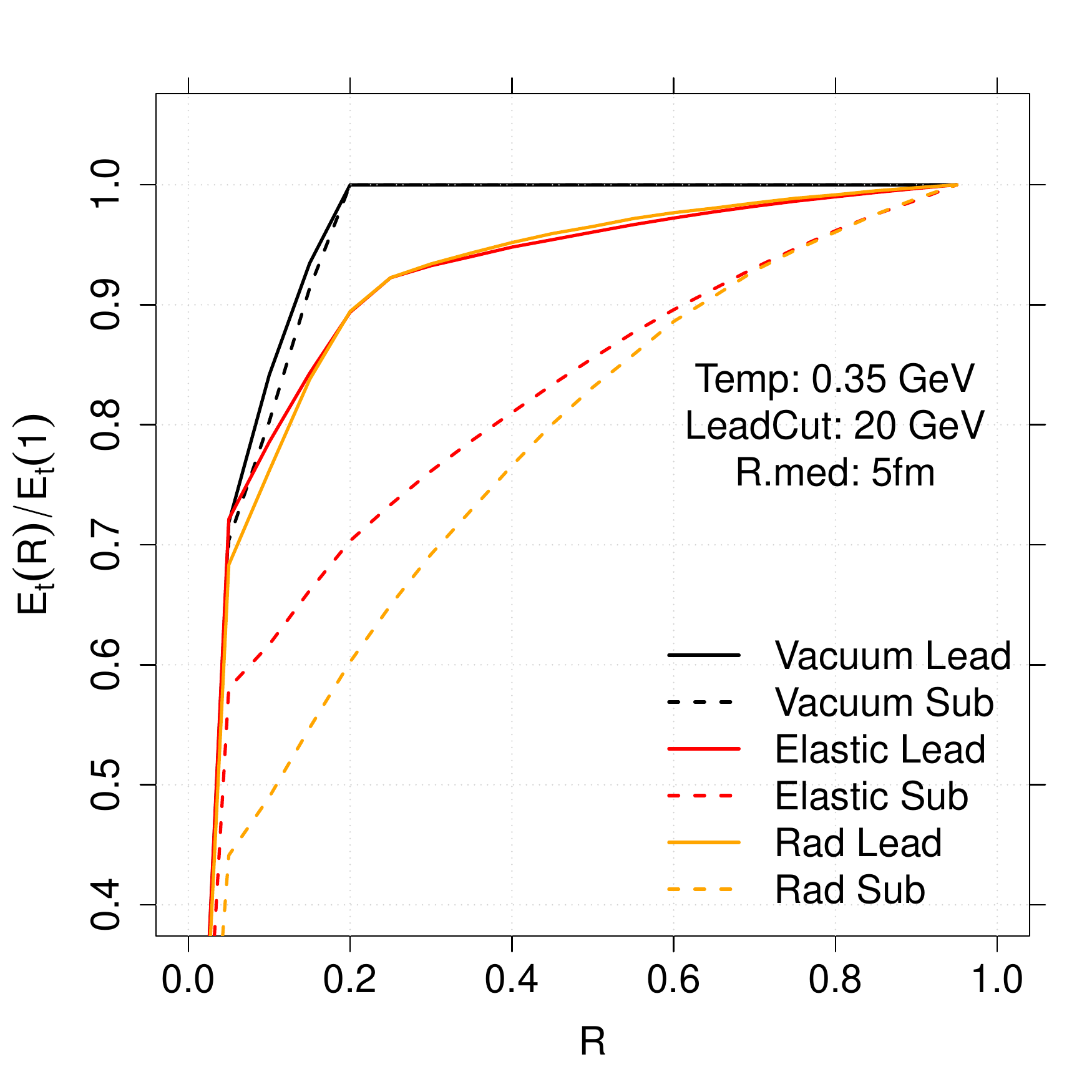}
  \caption{Variation of the jet radial profile at $T=250$~MeV (left) and $T=350$~MeV (right). Leading jet profiles are shown as solid lines and sub-leading jets are dashed, elastic only jets are red and radiative + elastic jets are orange. Note the strong softening of the sub-leading jet profiles and the clear separation of the elastic and radiative cases.}
  \label{fig-js-tvary}
\end{figure*}

\begin{figure*}[p]
  \includegraphics[width=0.45\textwidth]{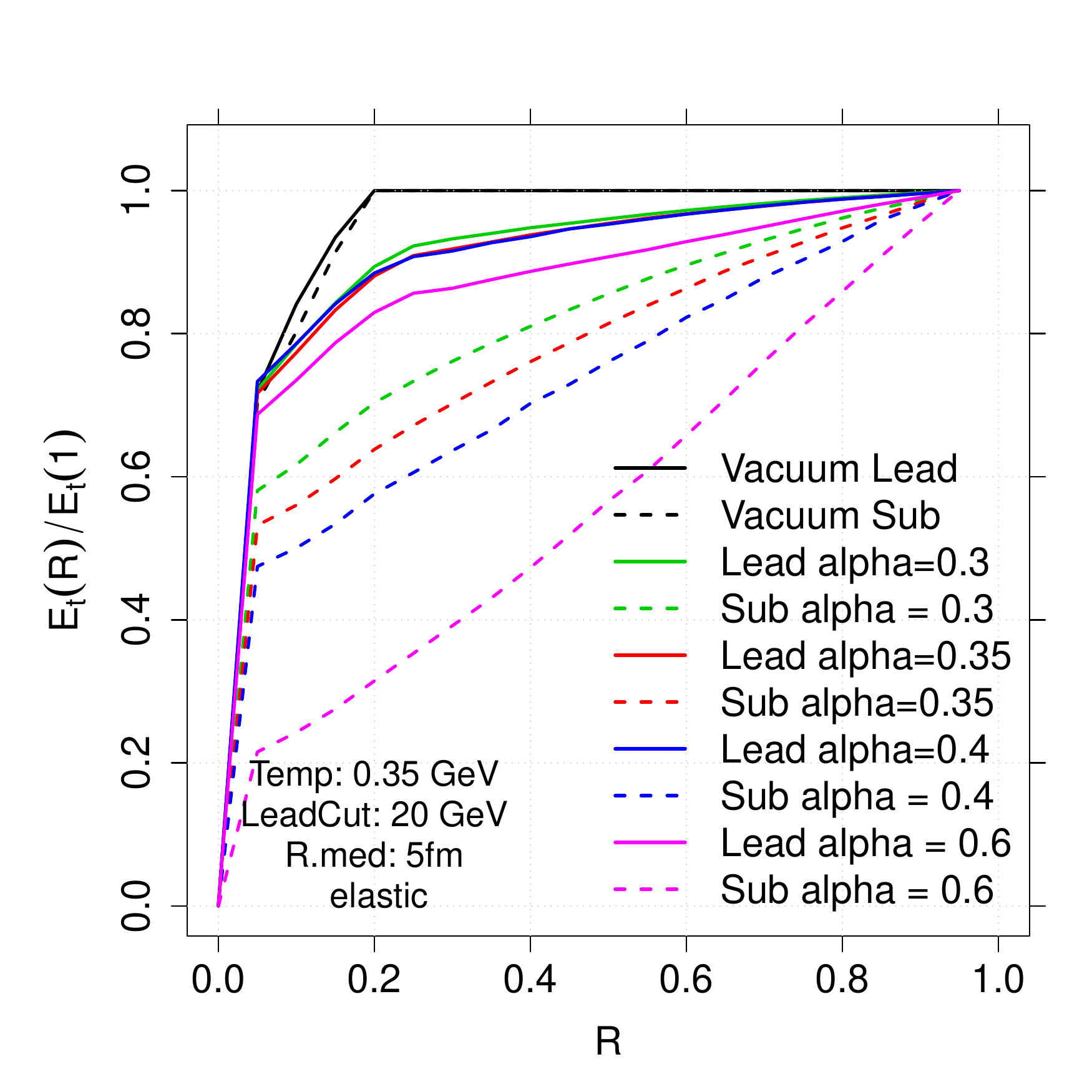}
  \includegraphics[width=0.45\textwidth]{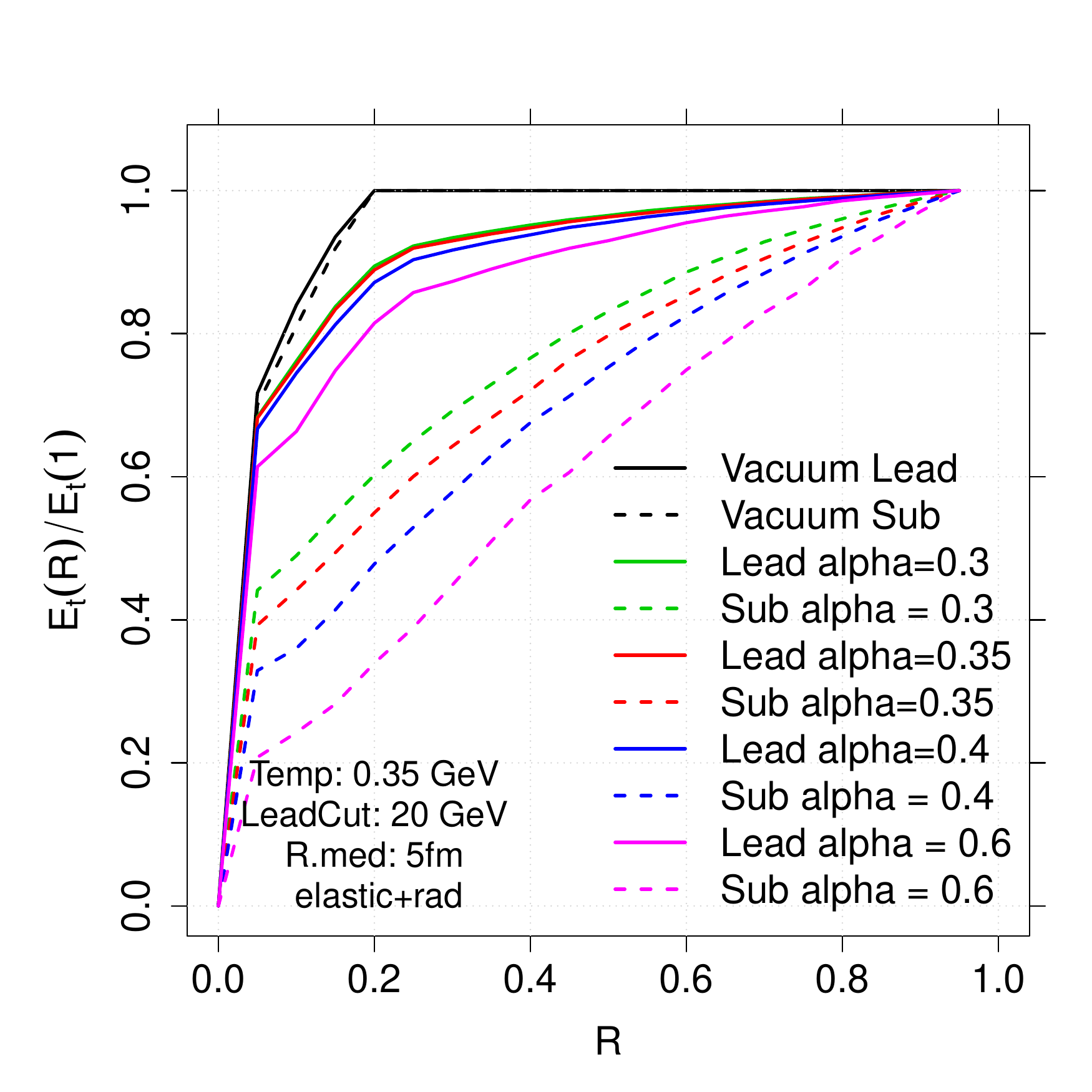}
  \caption{Variation of the jet radial profile at $T=350$~MeV. Leading jet profiles are shown as solid lines and sub-leading jets are dashed, elastic only jets are shown on the left and radiative + elastic jets are on the right. Increasing $\alpha$ strongly modifies the sub-leading jet profiles, note that the elastic (left) and radiative (right) sub-leading profiles take different shapes as $\alpha \to 0.6$.}
  \label{fig-js-alphavary}
\end{figure*}

Let us now examine the modification of the radial jet profile under the same set of factors. We define the radial jet profile as the ratio of jet energy reconstructed within a certain jet-cone radius $R$ relative to the reconstructed energy at $R=1$. This gives a normalized radial profile which is very sensitive to variations in the medium temperature, the jet interaction mechanism and the strong coupling. This observable could be experimentally obtained by a similar process of iterative jet reconstruction. In \figref{fig-js-tvary} we show the radial profile at two medium temperatures for both elastic and radiatively modified jets. The leading jet profiles (solid lines) are somewhat modified compared to the vacuum jets (black profiles), we see very little separation between the radiative and elastic only leading jet profiles. The sub-leading jet profiles are dramatically modified compared to the vacuum and leading jet profiles. At both temperatures the elastic and radiative sub-leading jet profiles clearly separate, the radiative sub-leading jets become broader and softer than the elastic only. Both sets of sub-leading jets become much broader and softer compared to the leading jets. This is in line with our observations of the energy loss distributions as shown in \figref{fig-de-joint}. In \figref{fig-js-alphavary} we show the jet profiles' sensitivity to the strong coupling. The leading jet profiles are somewhat modified at $\alpha_s=0.6$, the sub-leading jet profiles soften smoothly with increasing $\alpha_s$. It is interesting to note that the shape of these soft sub-leading profiles is different for elastic (left) and radiative (right) jets, the elastic profiles become increasingly convex as $\alpha_s \to 0.6$ while the radiative profiles remain concave. 


Finally we examine the partonic fragmentation distributions in the longitudinal and transverse variables $j_{t}$ and $z$. Examining the single inclusive jet distribution at RHIC scales $\sqrt{s} = 200$~GeV and  $E_t \sim 15 - 65$~GeV we find that the average partonic multiplicity in a vacuum PYTHIA jet is $\langle N_{parton} \rangle = 5.6$ which is roughly tripled during hadronization $\langle N_{hadron} \rangle = 14.6$, this hadronization process will conserve the total $E_t$ of the the jet but in so doing will redistribute the components of the momentum transverse to the jet axis. This means that the partonic and $j_t$ and $z$ distributions cannot be directly compared with experiment. What may be a surprise is the observed large peak in the $z$ distribution at 1 corresponding to jets with a single parton, this is seen in the vacuum and modified jet results. This peak is a consequence of the vacuum radiation evolution which is kinematically limited at RHIC regimes. We note that at LHC jet scales $E_t \sim 50-300$~GeV and $\sqrt{s} = 2.76$~TeV  $\langle N_{parton} \rangle = 19.5$ and $\langle N_{hadron} \rangle = 46.5$, at these scales the $z=1$ peak vanishes.

\begin{figure*}[p]
  \includegraphics[width=0.45\textwidth]{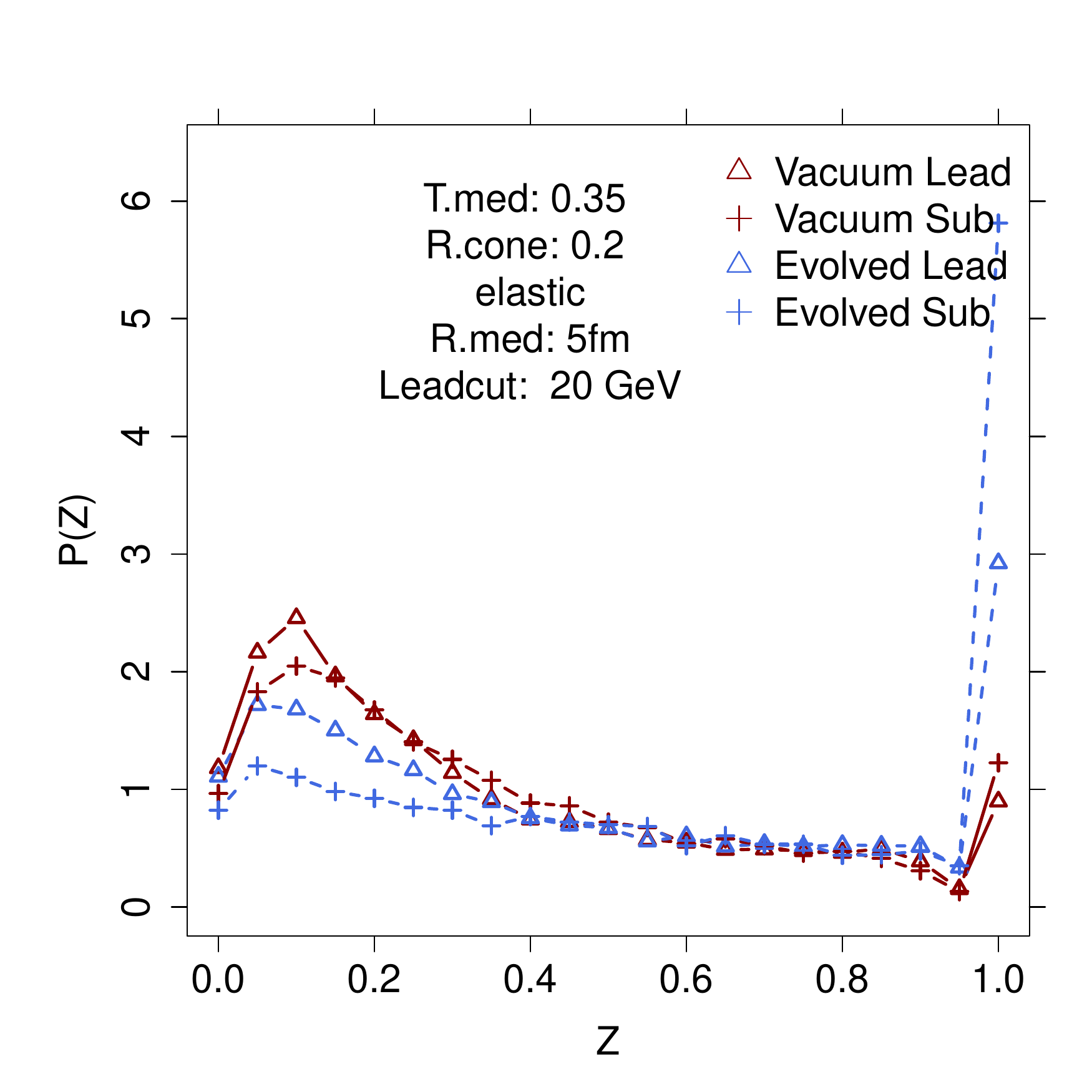}
  \includegraphics[width=0.45\textwidth]{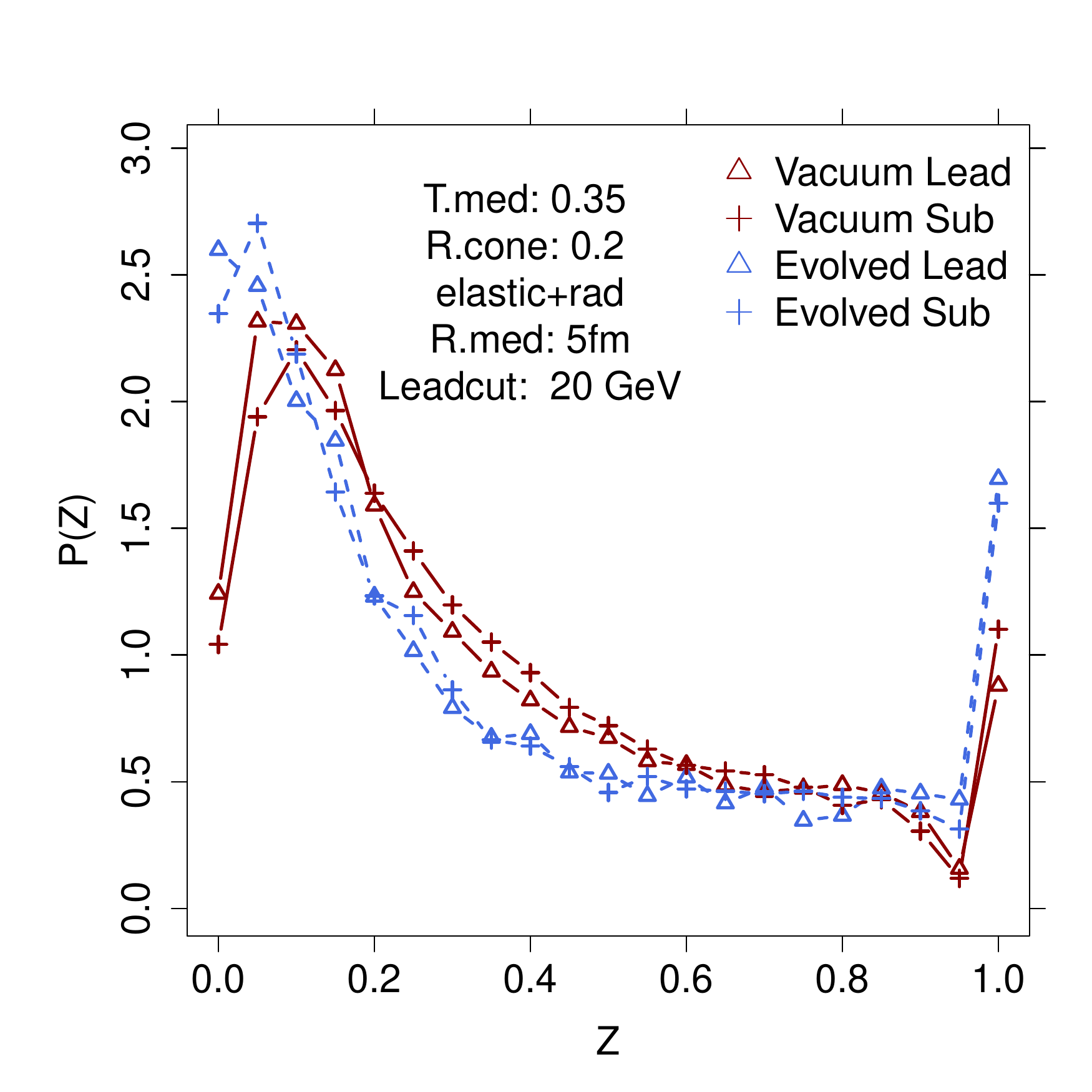}
  \caption{The $z$ distribution for partonic jets at $T=350$~MeV for jets reconstructed at $R=0.2$ for elastic jets (left) and radiative+elastic (right). Leading jets are shown as open triangles and  sub-leading jets are shown as crosses. The vacuum distributions are shown as red solid lines and the evolved distributions as blue dashed lines.}
  \label{fig-frag-z}
\end{figure*}

\begin{figure*}[p]
  \includegraphics[width=0.45\textwidth]{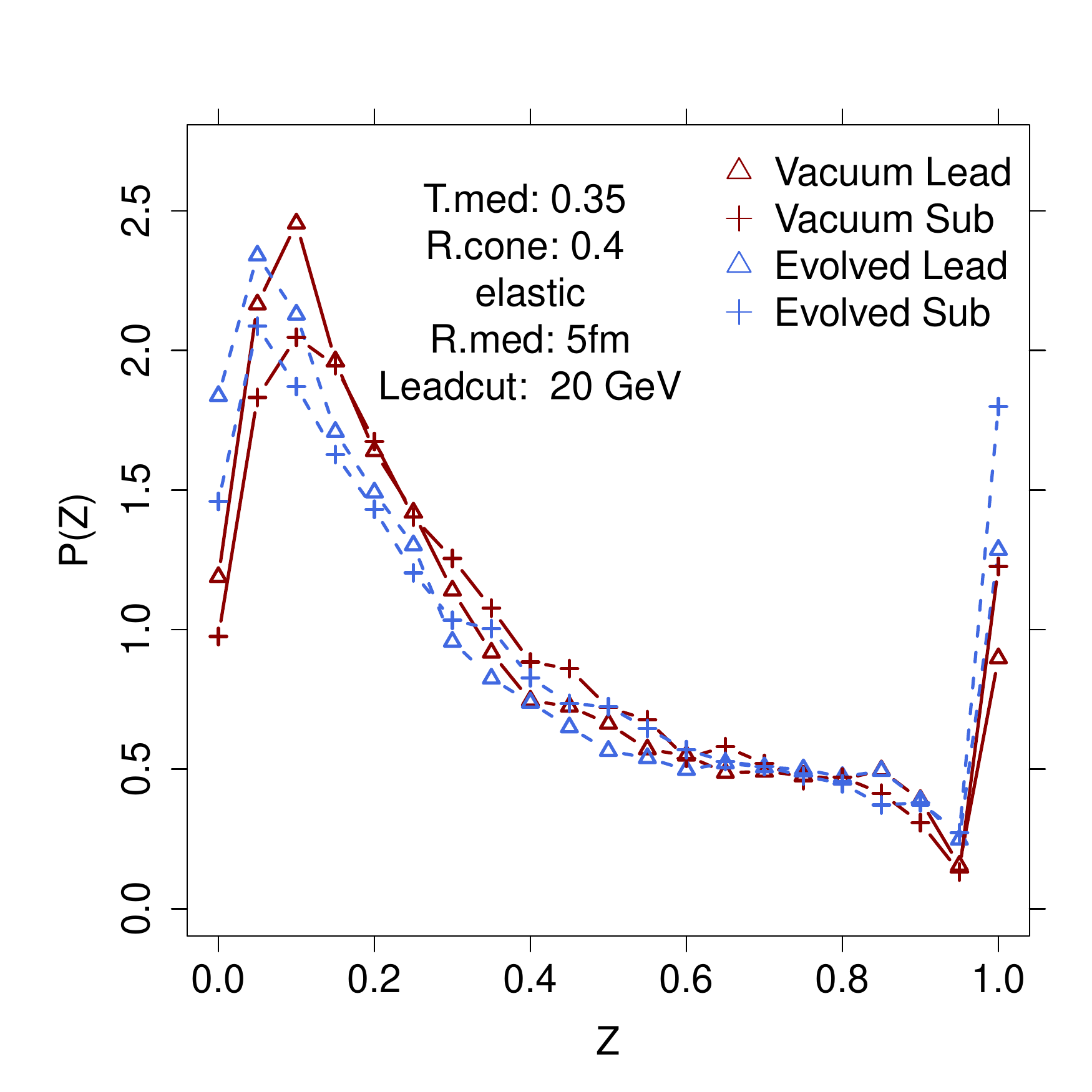}
  \includegraphics[width=0.45\textwidth]{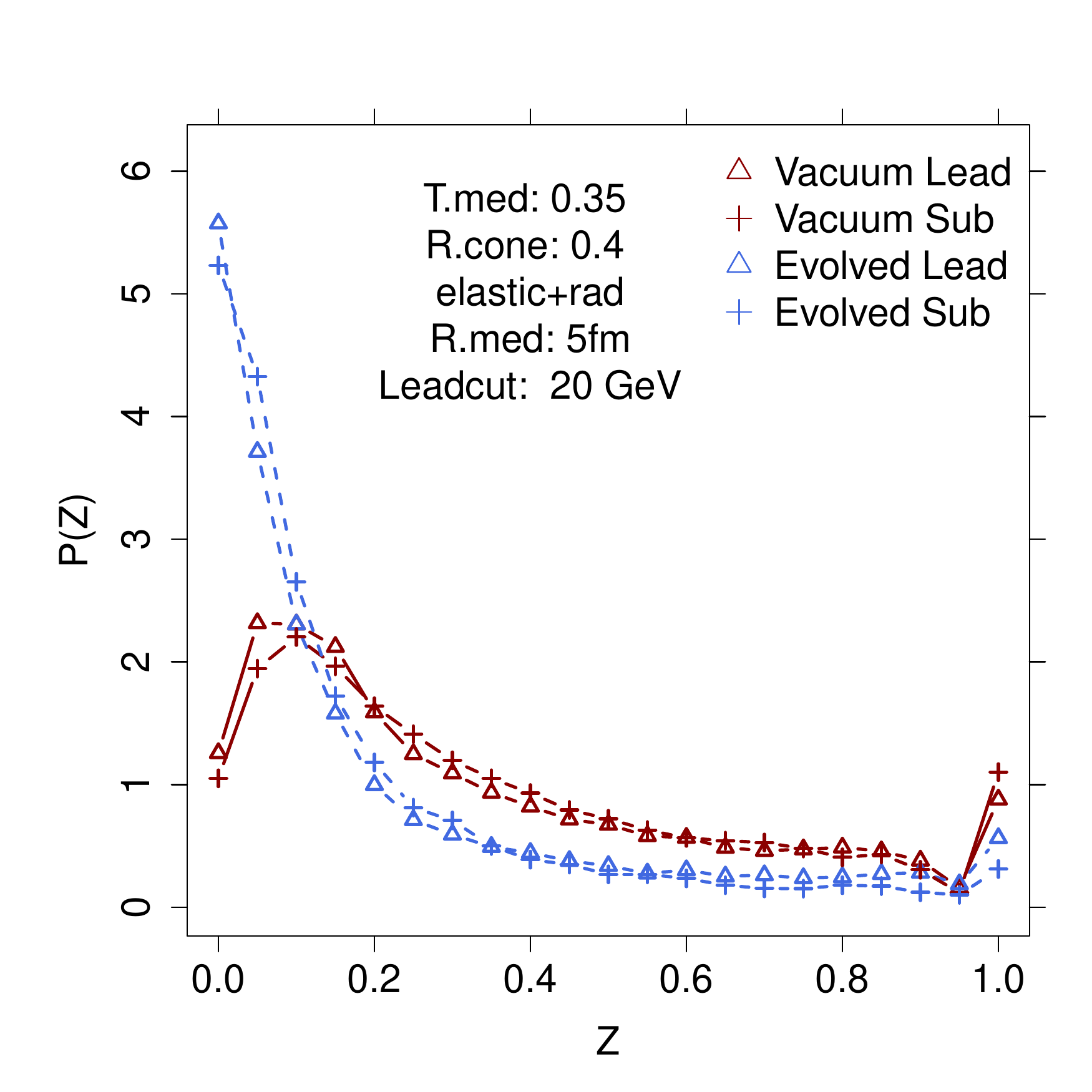}
  \caption{The $z$ distribution for partonic jets at $T=350$~MeV for jets reconstructed at $R=0.4$ for elastic jets (left) and radiative+elastic jets (right). Leading jets are shown as open triangles and  sub-leading jets are shown as crosses. The vacuum distributions are shown as red solid lines and the evolved distributions as blue dashed lines.}
  \label{fig-frag-z-wide}
\end{figure*}

\begin{figure*}[p]
  \includegraphics[width=0.45\textwidth]{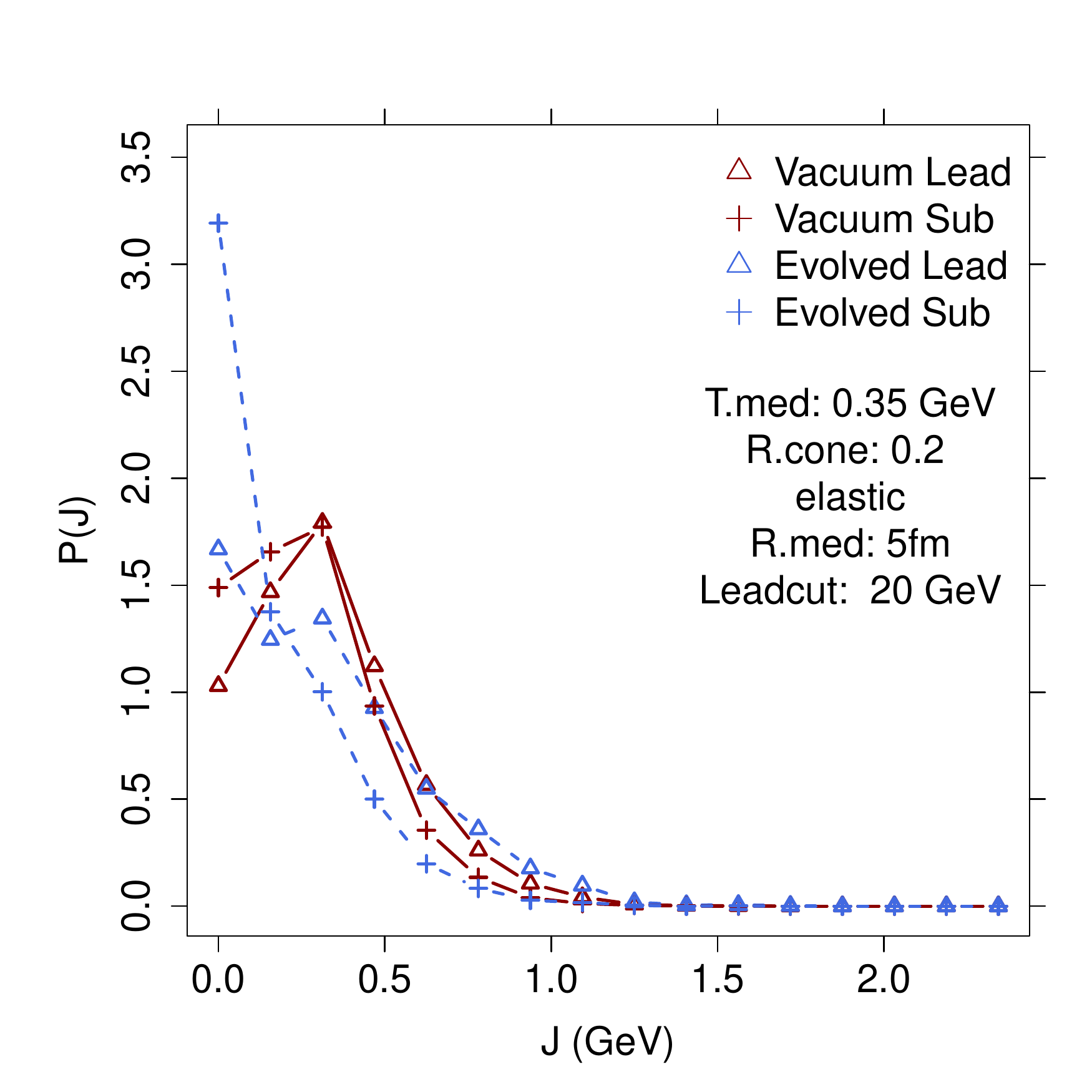}
  \includegraphics[width=0.45\textwidth]{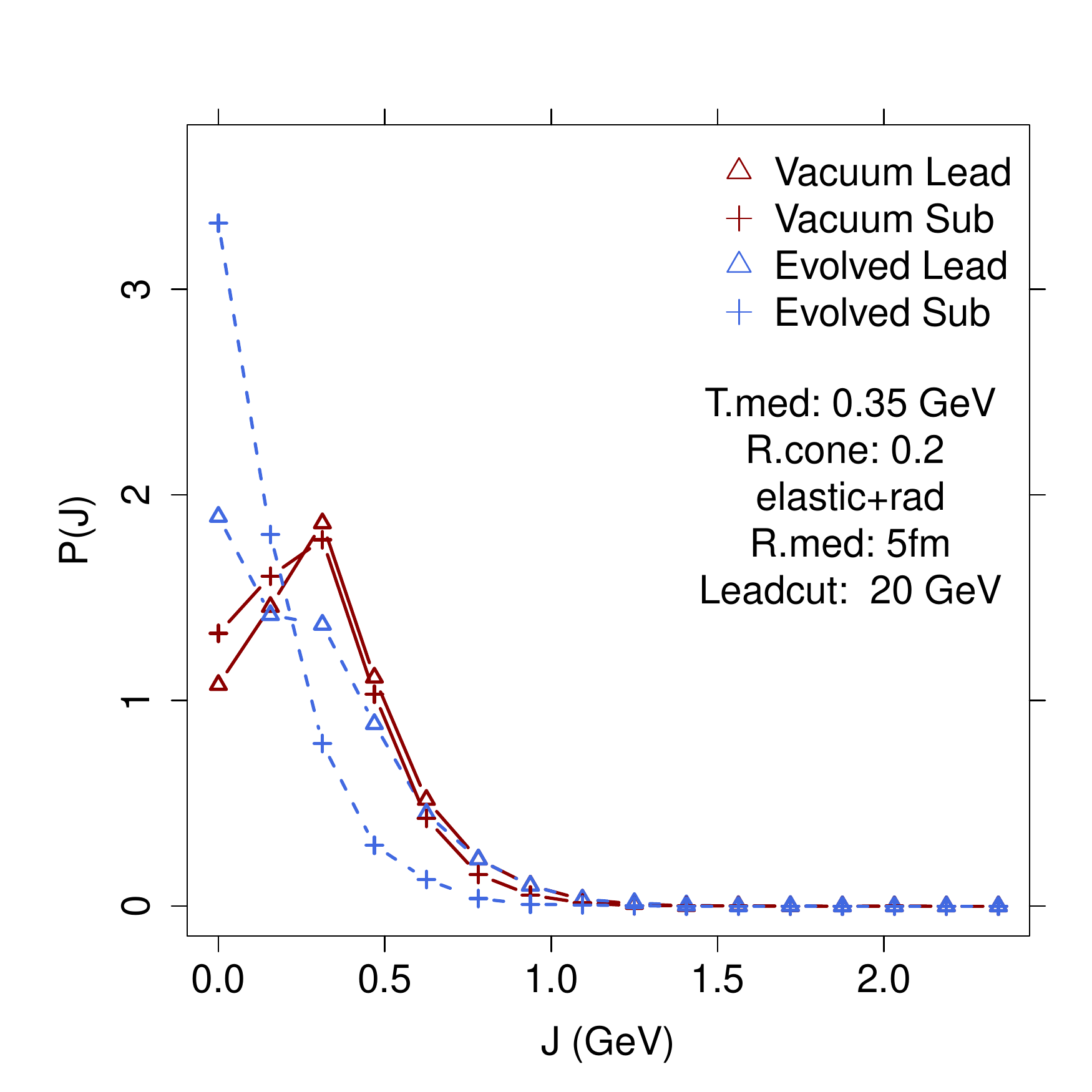}
  \caption{The $j_{t}$ distribution for partonic jets at $T=350$~MeV for jets at $R=0.2$, elastic only jets are shown on the left with elastic+radiative jets on the right. Leading jets are shown as open triangles and  sub-leading jets are shown as crosses. The vacuum distributions are shown as red solid lines and the evolved distributions as blue dashed lines.}
  \label{fig-frag-j}
\end{figure*}

In \figref{fig-frag-z} we show the VNI/BMS partonic $z = p_{t}/E_t \cos(\Delta r)$ distributions for elastic and radiative jets alongside the vacuum results. Note the strong peak at $z=1$ in both cases, this peak is actually enhanced by the evolution. In the elastic case this enhancement comes from a depletion of the low $z$ distribution. If a jet initially contains only one or two soft partons and a hard core the soft partons will most likely be scattered out of the jet cone. This leads to an enhancement of jets measured with $z=1$ and a depletion of the low end of the spectrum. The same process takes place in the radiative case, the right panel of \figref{fig-frag-z} but here the soft radiated partons tend to fill the small z distribution while the mid z range sees the most depletion. These middle $z$ partons interact with the medium producing a shower of soft radiation. The enhancement of the $z=1$ peak is smaller in the radiative case than in the elastic case as these hard core partons radiate and soften the jets. As the jet cone radius is increased the peak at $z=1$ is diminished as more of the soft scattered partons are recaptured by the jet definition this also enhances the peak at small $z$ in the radiative case, see \figref{fig-frag-z-wide}. The transverse $j_{t} = p_{t} \sin(\Delta r)$ distribution is shown in \figref{fig-frag-j} for elastic and radiative jets. The evolved $j_{t}$ distribution is softened in both cases as interactions with the medium and radiation add more soft partons to the jet. The radiative jets have a slightly steeper profile than the elastic only jets, radiated soft partons are more strongly confined to small $j_{t}$. 

\section{Conclusions}

We have examined the modification of RHIC scale dijets by a pQCD medium over a range of temperatures and coupling constants. We find that the dijet asymmetry is quite sensitive to the amount of medium encountered by the jets, and very sensitive to variations in $\hat{q}$. We do not observe a strong dependence of $A_{j}$ upon the jet interaction mechanism which is intriguing. The leading jets show a strong surface bias and lose very little energy making their $E_t$ a reasonable proxy for their initial energy. As such $A_j$ appears to give a reasonable measure of the modification of sub-leading jets at these scales.

The jet profiles we present show the strong modification of the sub-leading jets as a function of temperature and strong coupling. They also differentiate between interaction mechanisms, understanding the nature of the jet-medium interaction as well as its strength is of paramount import. Although these profiles may well be sensitive to the non-trivial hadronization process, we believe that they could provide a very informative window into the real nature of jet interactions with the QGP. 

The fragmentation distributions provide another interesting window into jet modification. The observed structures in $z$ reveal the importance of the interplay of vacuum radiation and the fragmentation process. Although hadronization will certainly change these distributions these results already demonstrate the importance of medium induced transverse diffusion in the jet modification process.

Dijets at RHIC scales are likely to be strongly modified by the presence of the deconfined QGP medium. The observables we have discussed are sensitive to many aspects of this modification and suggest that further jet measurements at RHIC will provide valuable insights into the nature of the QGP and into the applicability of pQCD jet suppression models .

\begin{acknowledgments}
We acknowledge support by DOE grants DE-FG02-05ER41367 and DE-SC0005396. This research was done using computing resources provided by OSG EngageVo funded by NSF award 075335. C.E.C-S would like to thank J.Nagle and S.A.Bass for many helpful discussions.
\end{acknowledgments}

\bibliographystyle{apsrev4-1}

\printfigures

\end{document}